\documentclass[final,1p,times]{elsarticle}

\usepackage{graphicx}
\usepackage{subfigure}
\usepackage{amssymb}
\usepackage{amsthm}
\usepackage{lineno}
\usepackage{booktabs}
\usepackage{threeparttable}
\usepackage{epstopdf}

\biboptions{sort&compress,square,comma}

\journal{Future Generation Computer System}

\begin{document}

\begin{frontmatter}



\title{Service Intelligence Oriented Distributed Data Stream Integration}

\author{Feng-Lin Li, Chi-Hung Chi, Yue Wang, Cong Liu}
\ead{lifl08@mails.tsinghua.edu.cn}
\address{School of Software, Tsinghua University, Beijing, China, 100084}

\begin{abstract}
Software as a service (SaaS) has recently enjoyed much attention as it makes the use of software more convenient and cost-effective. At the same time, the arising of users' expectation for high quality service such as real-time information or functionality provisioning brings about new challenges: to satisfy such (near) real-time requirements, real-time monitoring and effective processing of streaming data is necessary. However, due to the composition structure and multi-instance property of services, service data streams are often distributed, hard to synchronize and aggregate. We tackles these challenges by (1) proposing systematic associate strategies for relating distributed data; (2) introducing a new small window array mechanism for aggregating distributed data; (3) setting window parameters based on the cumulative distribution function (CDF) method; and (4) modeling streaming operators with queuing models for performance evaluation and prediction. Experiments show that our approach has good accuracy, completeness
and acceptable performance measurement in processing distributed service data streams.
\end{abstract}


\end{frontmatter}


\section{Introduction}
Along with the movement of information technology, the majority of
software nowadays are presented in the form of services over
Internet. In this way, most of the implementation details can be
well hidden behind interfaces which are simple and user-friendly to
users, making software easier to use in spite of its increasingly
complicated internal building blocks. \textbf{\emph{Service}} is a
set of functionalities which can perform a specified task or supply
corresponding information with the addition of its deployment and
runtime environment \cite{elsarticle-num:Erl_SOAPrinciples,
elsarticle-num:jones_toward_2005}. Providing software as a service
emerges as a novel service delivery model in recent years, in which
people use software on-demand instead of purchasing the entire
software package, makes the using of software more convenient and
cost-effective \cite{elsarticle-num:software_service} and has
obtained wide attention from both academic and industrial circles.

The general trend indirectly reflects the customers' interesting and
potential demand for satisfactory services. As time moves on,
people's expectation for high service quality and intelligence
brings novel requirements and challenges to services. Though
undeniablly current services and traditional off-line analysis have
done a pretty good job to cater to customers' demand, there are
still some aspects - say real-time functionality, information,
service intelligence (dynamic scaling) - that they don't perform
well in.

Generally in services nowadays, static data is operated on to
produce desired values which may become invalid when it comes to
time varying data. Supposing that a driver want to find the nearest
available parking space, the service may have to gather the parking
information of adjacent parking areas timely and provide reasonable
outcomes after instant computation.

Another important aspect of service intelligence is service
management. A customer may expect the service he/she has subscribed to
could keep working with high performance. In other
words, a customer may hope that a service could deliver a valid
response within 3 seconds even at the visit peak.

As is known, traditional off-line analysis may provide some
intelligence - say service requirement, recommendation - through
data mining or analysis. However, they won't be able to satisfy the
real-time requirement because of the time limitation. In order to
provide real-time information or guarantee high service quality,
real-time data collecting and processing is the essence.

Data usually comes from different sources in practice. As for a
real-time functional service, it may need to gather information from
different terminals like sensor, agent and smart phone to conduct
computation. On the other hand, the composite structure and
multi-copy property of service also contribute to the data
distributivity. Usually for convenience and reusing, a service is
composed of sub-services, each of which provides specified
functionalities and may be deployed on different nodes. Also, a
service may have multiple copies running on the same platform in
order to respond to to the users' requests timely during peak time.
Due to the distributivity of service, multiple data streams will be
generated on sequential requests (each node will generate an event
stream in which a tuple stands for an invocation). In order to
provide relevant function or intelligence, reasonable preparation
and integration over the multi-streams should be taken as the
fundamental step.

In fact, real-time information or functionality service is a kind of
service intelligence - quality requirement. In this paper, we mainly
pay attention to distributed data integration towards upper service
intelligence provision. The remainder of this paper is structured as
follows: Sec. 2 presented the relevant work; Sec. 3 analyze
significant problems of distributed data stream information
integration in service environment; Sec. 4 proposes different
associate strategies to address the identifier problem; Sec. 5
introduce the small window array mechanism and employs the
cumulative distribution function approach to configure stream system
parameters for better integration completeness; Sec. 6 builds
reasonable queuing model for typical stream operators involved in
this paper to predict performance indicators; finally, we draw our
conclusion and give some future research issues in section 7.

\section{Related Work}

\subsection{Service Intelligence}\label{Sec.SI}

Except for the functionality, service intelligence mainly includes
requirement, trust, recommendation, management and so on. In order
to provide customized service, the key point is to obtain a deep
understanding of service quality requirement. SiMing Li
\cite{elsarticle-num:li_adaptive_2008} proposed a framework of
acquiring service non-functional requirement through user-oriented
behavior information and service-oriented runtime as well as static
information (i.e. service structure, SLA). An example is that some
people put great attention on quality while others only care about
the price. In addition to the traditional quality requirement like
reliability and availability, real-time information or functionality
turns to be another kind of significant quality requirement and
relies heavily on the underlying real-time data gathering and
processing. For instance, ticket information inquiry and real-time
stock market analysis might fall into this category.

Recommendation is of great significance to both service providers
and customers. Based on the subjective and objective information,
providers can know well about the Service Level Agreement (SLA)
compliance (the conformity degree between practical measurements and
guaranteed service quality in
SLA)\cite{elsarticle-num:keller_defining_SLA_2002,
elsarticle-num:li_adaptive_2008}, customers' interests and hence
make reasonable decisions. At present, customer-oriented
recommendation mainly hires the collaborative filtering method and
falls into two classes: user- and content-based recommendation
\cite{elsarticle-num:sarwar_item-based_2001}. The customized
recommendation of blog, commodity and trip modes that press close to
users' interest can bring customers better experience.A route
selection service which can analyze real-time route information and
recommend an uncrowded route will be a good case in point. It could
become more wonderful if user preference is also taken into
consideration.

In order to provide customized service for individuals, service
providers need to gather as much user-generated content
\cite{elsarticle-num:burtch_user-generated_content} as possible
instead of asking them to fill in an electronic form explicitly. The
user-generated implicit data includes the query keywords in a search
engine, browser history on web-sites, purchasing behavior, etc.
Comparing with the explicit information like manual feedbacks and
reviews, implicit data generated by users unconsciously is of great
value in acquiring users' interests and preference, which can
contribute a lot to service requirement analysis, reputation and
recommendation.

Service management includes SLA compliance, problem diagnosis,
dynamic scaling etc. SLA compliance indicates the conformity between
the actual runtime measurement and the guaranteed quality. Problem
diagnosis is of great importance when a service takes a long time to
response or even fails to respond. Problems have to be located
accurately and got fixed as soon as possible in order to minimize
the negative effects. Last but not least, dynamic scaling plays a
key role in the maintenance of service scalability. As is well
known, the number of visits varies over time, and thus dynamic
scaling mechanism of service management should be able to perceive
the changes and take appropriate countermeasures.

From the discussion above, we can get a clear picture of real-time
service intelligence that rely on time varying data. A plenty of
research have been done towards service intelligence like trust,
requirement and recommendation on the basis of static data.
Researchers usually assume the required data to be ready when
verifying their models or algorithms. In fact, when it comes to
real-time scenario, how to gather distributed streaming data and
prepare them well for kinds of superior intelligence models is a
Gordian knot.

\subsection{Data Stream}\label{subsec.stream}
A data stream is an infinite sequence of tuple
\cite{elsarticle-num:schmidt_quality--service-aware_2007}. Similar
with the data table of database system, a data stream has an
appropriate schema that is defined by a collection of attributes
with specified data types. Generally, a stream has a timestamp
located in the first column which records the tuple's generation
time. According to the interval time between tuples, stream falls
into two classes: \emph{uniform} and \emph{random}. Intuitively, a
data stream can be characterized by a two-dimension array. Each row
in a stream represents a tuple while each column stands for a
specific attribute.

Comparing with static data in traditional database system, streaming
data is dynamic and transient. It can be processed only once at each
operator and then sent away or discarded. Because of the
time-varying stream rate and limited processing capacity, sometime
the stream engine has to drop some tuples or use synopsis to obtain
approximate results. In addition, streaming data is updated in the
form of appending instead of modifying and its model (schema) can be
modified by operators dynamically during processing.

Another significant difference between data stream and database is
the {\bf continuous query}. A database will return back static and
eligible data on a query and the result won't change even if updates
are delivered soon afterwards. Being different from the database, a
query in data stream system is installed permanently before it is
canceled explicitly by the user. The query is triggered by the
arrival of each incoming tuple. Apparently, the query result is
updated continuously and becomes a stream
\cite{elsarticle-num:koudas_data_2003,
elsarticle-num:babcock_models_2002,
elsarticle-num:golab_issues_2003}.

Numerous studies have been done in different aspects of
distributed data stream research. B. Babcock etc.
\cite{elsarticle-num:babcock_distributed_2003} proposed an
approximate but effective Top-K algorithm to calculate the K most
popular pages with prerequisite of minimizing data communication
between computation and coordinate nodes. C. Olston etc.
\cite{elsarticle-num:olston_adaptive_2003} considered an adaptive
filter algorithm which adjusts the tolerance according to users'
requirement in order to reduce redundant information under
multi-user query scenario. Samuel Madden etc.
\cite{elsarticle-num:madden_fjording_2002} put forward a Fjord
framework which joins sensor streaming data with traditional static
data. It is pretty useful to calculate SLA compliance by combining
dynamic runtime information with static service or SLA information.

Current work in data stream has made excellent progress and
developed fairly mature prototypes, such as BOREALIS
\cite{elsarticle-num:borealis, elsarticle-num:abadi_design_2005}
system of MIT, Brown and Brandy University, Telegraph-CQ
\cite{elsarticle-num:TelegraphCQ,
elsarticle-num:TelegraphCQ_Uncertain_World,
elsarticle-num:TelegraphCQ_Adaptive_Query} system of UC Berkeley,
STREAM \cite{elsarticle-num:Stanford_STREAM_Main,
elsarticle-num:Stanford_DSMS} of Stanford University. Among them,
Telegraph-CQ and STREAM employ Continuous Query Language (CQL)
\cite{elsarticle-num:babcock_models_2002,
elsarticle-num:golab_issues_2003} as the query language while
BOREALIS use XML. All the above three data stream system supports
sliding window but STREAM doesn't support the regulation of advance
step which is fixed to 1 by default. They can accept multiple
streams as inputs from network or file and deliver query results to
multiple users at the same time.

We choose BOREALIS stream system as our research and experiment
platform because its distributivity is really suitable for our
service environment. The key concept of BOREALIS is operator box
\cite{elsarticle-num:Borealis_Guide}. The operators fall into two
classes: stateless and stateful. Typical stateless operators contain
MAP, FILTER and UNION; operators such as JOIN, SORT and AGGREGATE
are stateful. Another significant property of BOREALIS is the good
extendability that stems from its inheritance mechanism with which
users can define and implement custom operators like clustering by
inheriting the \textbf{\emph{QBox}} super class and overwrite the
\textbf{\emph{run\_impl}} function. In this way we obtained our own
sliding window aggregate operator.

\subsection{Queuing Theory}
Queuing theory is a mathematical study for waiting queue and
contains three key stages: input/arrival process, waiting for
processing and receiving service in the system. In view of the
Markov process's steady-state balance equation and the Little rules
\cite{elsarticle-num:gross_fundamentals_2008}, queuing model can be
used to predict kinds of indicators like waiting queue length
(customers in the waiting line), queue length(all the customer in
the system), waiting time, residence time, and the probability of
system is busy or the probability a tuple will be dropped.

Queuing model can be presented as $A/B/C/D/E/F$ via Kendall symbols
\cite{elsarticle-num:gross_fundamentals_2008,
elsarticle-num:osogami_analysis_2005}. $'A'$ represents the input
interval distribution which usually includes poison distribution
($M$), K-Erlang distribution ($Ek$) and General distribution ($G$);
 $'B'$ stands for the service time distribution and has similar types
of distribution with input process; $'C'$ means the number of
service desk; $'D'$ is the buffer length; $'E'$ indicates whether
the data source is infinite or not; $'F'$ is the service discipline,
like first come first served (FCFS, default discipline), last come
first served (LCFS).

The most common distribution for input process and service time is
{\emph{negative exponential distribution}}( also known as
{\emph{poisson distribution}}, $M$). This kind of queuing model
contains $M/M/1$ and $M/M/C$, which means a queuing model with both
poisson input process and service time distribution, 1 or $C$
server(s). The model can be extended to $G/M/1$ or $M/G/1$ if the
input process or service distribution becomes general (e.g. K-Erlang
distribution, deterministic and poisson). The normal \emph{Poisson}
and \emph{Erlang} distributions have apparent limitation in
practical applications because of their relative simplicity.
Therefore we mainly employ Hyper-Erlang
\cite{elsarticle-num:Erlang_distribution,
elsarticle-num:xu_performance_2008} and Phase-Type (PH) distribution
\cite{elsarticle-num:phase-type_distribution} to fit general
distribution from real streaming data.

\textbf{\emph{Erlang distribution}} is of significant importance to
the Expectation Maximization \cite{elsarticle-num:PH_EM} approach.
Its \textbf{\emph{probability density function (PDF, $f$) }}and
\textbf{\emph{cumulative distribution function (CDF, $F$)}} are
shown in formula (\ref{Formula.01}).

\begin{equation}\label{Formula.01}
f(x,k,\lambda) =
\frac{{\lambda}^{k}x^{k-1}e^{-{\lambda}x}}{\prod\limits_{i=1}^{k-1}i},(x,\lambda
\geqslant 0) \qquad F(x,k,\lambda) = 1 - {\sum_{n=0}^{k-1}
\frac{e^{-{\lambda}x}({\lambda}x)^{n}}{\prod\limits_{i=1}^{n}i}}
\end{equation}

\textbf{\emph{Hyper-Erlang Distribution}} is a mixture of $m$
branches of Erlang distribution. The $i$-th Erlang branch has an
initial probability $\alpha_{i}$ and the sum of all $\alpha_{i}$
equals to 1. Each branch of Erlang distribution has a parameter pair
$E_{i}(\lambda,K)$ in which $\lambda$ is the rate and $K$ represents
the scale(number of phases). The PDF and CDF of Hyper-Erlang
Distribution can be expressed as formula (\ref{Formula.02}) based on
the formula (\ref{Formula.01})
\cite{elsarticle-num:Erlang_distribution,
elsarticle-num:xu_performance_2008}.

\begin{small}
\begin{equation}\label{Formula.02}
f_{HErD}= \sum_{i=1}^{m} \alpha_{i}f_{i} = \sum_{i=1}^{m}
\alpha_{i}\frac{{\lambda_{i}}^{k_{i}}x^{k_{i}-1}e^{-{\lambda_{i}}x}}{\prod\limits_{i=1}^{k_{i}-1}i},(x,\lambda_{i}
\geqslant 0) \quad F_{HErD}= \sum_{i=1}^{m} \alpha_{i}F_{i} =
\sum_{i=1}^{m} \alpha_{i}({1 - {\sum_{n=0}^{k_{i}-1}
\frac{e^{-{\lambda_{i}}x}({\lambda_{i}}x)^{n}}{\prod\limits_{j=1}^{n}j}}})
\end{equation}
\end{small}

A \textbf{\emph{PH distribution}} with parameter $(\alpha, T)$,
$PH(\alpha, T)$, is the distribution of the time until absorption
into state 0 in a Markov chain on the states $\{1, \ldots, n, 0\}$
with initial probability vector $(\alpha, 1 - \alpha \times
\textbf{1})$ and and infinitesimal generator:
\begin{equation}\label{Formula.13}
    Q = \left(
          \begin{array}{cc}
            T & T^{0} \\
            \textbf{0} & 0 \\
          \end{array}
        \right)
\end{equation}
where $\alpha$ is a $1 \times n$ vector, $T$ is a $n \times n $
matrix and $T^{0} = -T \times \textbf{1}$, here $\textbf{1}$
represents an $n \times 1$ vector with every element being 1.

Hyper-Erlang Distribution and PH distribution will play significant
role in our queuing model. In addition, the service model of queuing
system falls into two classes: single and batch service
\cite{elsarticle-num:gupta_analysis_2001,
elsarticle-num:banik_finite-buffer_2009,
elsarticle-num:laxmi_finite-buffer_1999}. In simple words, the bach
service model is of great importance for stateful stream operators
with window mechanism. For modeling details, please refer to section
\ref{Sec.Queuing}.

\section{Problem Analysis}

\subsection{Service Intelligence Environment}
\subsubsection{Data Stream based Service Intelligence System}
Various services can be deployed on a service platform, including
utility services such as identification, security, and monitoring
services, and functional services like search, shopping and
consulting services. Taking book search services for example, here
we present a brief explanation for service recommendation. Assume a
book search service set $\{S_1,S_2,\ldots,S_n\}$ with similar
functionalities are deployed on a service platform. Facing so many
similar services, which one should a new user choose when he/she has
just got on the platform? If records are kept about recent behavior
of each service on the platform, like whether the service quality
conforms to the promise, how much percentage of users are using it,
and to what extent users are satisfied with results provided by the
service, such information can be fed to some trust or recommendation
algorithms for further computation. Having the necessary
information, we can then recommend the most popular search services
to users according to some sorting rules, thus providing better
service quality.

At some degree, those varieties of intelligence models such as
requirements, trust, reputation share not only some common
procedures but also mutual data basis. On one hand, given a
computation procedure library constituted of common procedures,
different intelligence models can be conveniently constructed by
combining procedures. When we are in need of a new model,
combination of current available procedures are first examined. New
computation procedure will be introduced into the layer if none of
the combinations works. On the other hand, the above
service-oriented objective and user-oriented subjective information
constitutes the data basis of service intelligence algorithms.

Based on these ideas, we propose the concepts of the algorithm layer
and the data layer. The algorithm layer is the collection of
different formulas and algorithms like clustering algorithms,
distance-based algorithms, etc. The data layer collects data from
the service platform and performs real-time processing to give out
the uniform data distribution, which then is shared by these two
layers as the basis of communication and interaction.

The algorithm layer and the data layer are the core components of
the Data Stream based Service Intelligence System(DSSIS), which aims
at extracting common features such as requirement, recommendation
and management and building a fundamental system for service
intelligence with uniform data distribution and common algorithms.
The framework of DSSIS is composed of four layers which are the
model layer, the algorithm layer, the data distribution (the
knowledge layer) and the data layer respectively.

\begin{figure}
  \centering
  \includegraphics[width=0.80\textwidth]{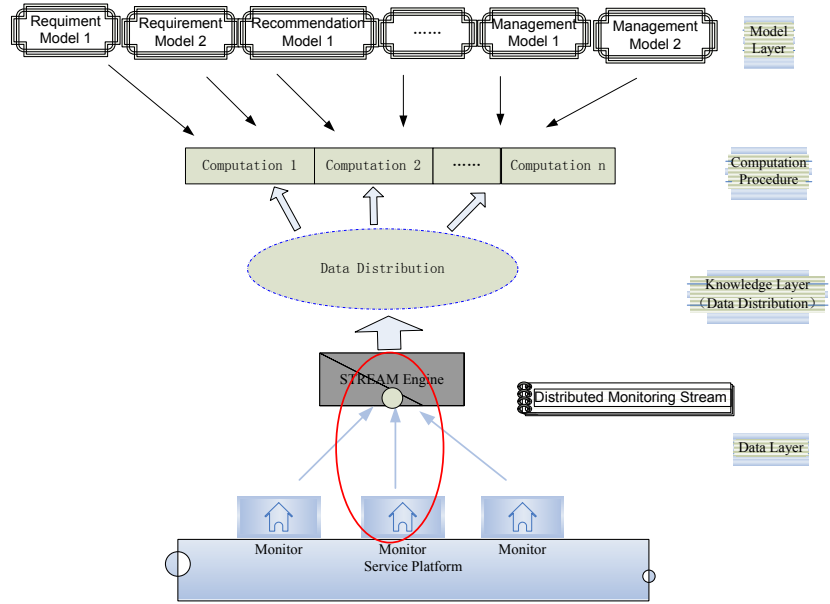}\\
  \caption{Data Stream based Service Intelligence System
  Framework}\label{Fig.01}
\end{figure}

\begin{itemize}
  \item \textbf{\emph{The model Layer}} is on the top of DSSIS in which models
  differ from application to application. Representative models includes recommendation,
  user classification,
  etc.
  \item \textbf{\emph{The algorithm layer}} is a collection of algorithms designed
for obtaining information for the upper models. A large number of
algorithm classes are located in this layer, each of which includes
multiple algorithms. Typical classes includes (1) Clustering, (2)
Averaging, (3) Time-decaying.
  \item \textbf{\emph{The knowledge layer}} is a uniform data distribution which
  updates with certain frequency. A data distribution is the summary of
  runtime service measurement, user behavior and feedbacks during a period.
  In addition, it can be persisted and the evolving of data distribution
  can be valuable for prediction as time goes on. The data distribution
  is structured to meet the input requirement of upper algorithms.
  \item \textbf{\emph{The data layer}} collects streaming data from different
service nodes and performs appropriate preparation and integration.
The data comes from monitoring services deployed on the platform and
transfers between nodes as data streams. Due to the composite
structure and multi-copy of services, the integration of distributed
data streams brings novel challenges.
\end{itemize}

To obtain model knowledge, the model layer may require different
classes of algorithms; The structure of the data distribution in the
knowledge layer needs to meet the input standard of the algorithm
layer; Based on the algorithms' requirements, the data layer
integrates data from distributed streaming data and produces a
uniform data distribution. As is seen, these four layers are
correlated, modification in one layer may have effects on another
one. Our study in this paper mainly focuses on the data integration
in the fundamental data layer.

\subsubsection{Service Property}
Service is usually deployed on a platform and is composed of other
services because of complicated business logic. Sometimes it also
has multiple copies running on the platform in order to guarantee
the scalability. We will give a brief introduction of service
properties at first.

\begin{itemize}

\item{\textbf{Service}}

{Service} includes atomic and composite service
\cite{elsarticle-num:SOA_Reliability}. An {\bf \emph{atomic
service}} can either offer utility independently or can be a
component of composite services. A {\bf \emph{composite service}} is
composed of multiple services, each of which may be atomic or
composite. Accordingly, an {\bf \emph{independent service}} is an
atomic service which offer utility independently or a final
composite service. The components (atomic or composite service) of
an independent service are called {\bf \emph{sub-services}}.The
number of sub-services is called {\bf \emph{service degree}}.

From the definition above, we come to know that a composite service
is defined iteratively. The root \-- which is called {\bf \emph{head
service}} \-- is used as the representative of the entire service.
Meanwhile, head service is also a sub-service. Sub-services except
for the head are called {\bf \emph{subordinate services}}. When a
user invokes the head of a composite service, its response will
guide the user to the subordinate services.

\item{\textbf{Invocation}}

An invocation is a request sent from the client to a certain service. A
{\bf \emph{primary invocation}} is an invocation of an independent
atomic service or the head service of a composite service. A {\bf
\emph{subordinate invocation}} is the invocation of the subordinate
service which is caused by the primary invocation. An invocation of
atomic service is primary when it is independent while subordinate
if it acts as a service component.

\item{\textbf{Instance}}

An instance includes at least one primary invocation and zero or
more subordinate invocations of the same service. The number of invocations in an
instance is called the {\bf \emph{instance degree}}.
\end{itemize}

\begin{figure}
\begin{center}
  \includegraphics[width=0.90\textwidth]{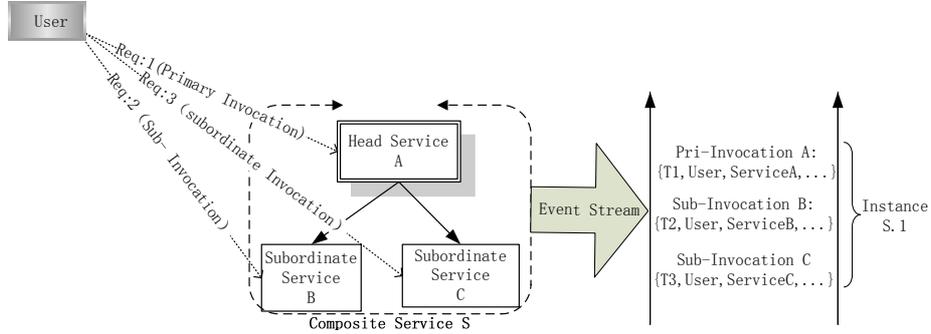}\\
  \caption{Relations of Service, Invocation and Instance}\label{Fig.02}
  \end{center}
\end{figure}

The relations among service, invocation and instance are illustrated
in Figure \ref{Fig.02}. We assume that each service, no matter
atomic or composite, should be invoked by the client directly. The
head service will not invoke the subordinate services on its own but
its response will guide the client to the remaining subordinate
services. It looks a bit like a HTML file of the web page.

\subsection{Case Study}\label{Sec.Case}
Through a specific case of service recommendation model, this
section explains how to make use of the stream integration result to
support upper intelligence model in DSSIS, presenting the important
role that the data layer plays in the whole system.

\begin{figure}
\begin{center}
  \includegraphics[width=0.35\textwidth]{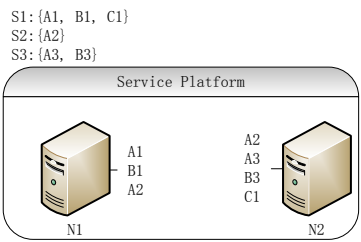}
  \caption{Service deployment illustrations}\label{Fig.03}
\end{center}
\end{figure}

Assuming there are two computation nodes $N_1, N_2$ and three
services $S_1, S_2, S_3$. $S_1$ includes atomic service $A_1$ and
subordinate services $B_1$, $C_1$(which indicates a service degree
of 3); $S_3$ includes $A_3$ and $B_3$; $S_2$ is an independent
atomic service. We also suppose that $A_1, B_1, A_2$ are deployed on
node $N_1$ while $A_2, A_3, B_3, C_1$ are hosted on $N_2$. The
deployment diagram is illustrated in figure \ref{Fig.03}. Please
note that service $S_1$ is deployed separately and $S_2$ has two
copies running on the platform. Every time a service is invoked, the
information about the invocation will be collected, producing
distributed monitoring data streams. We define a schema for the
invocation stream: $\{Timestamp, UserId, ServiceId, ResponseTime\}$.
An invocation tuple implies a user invoked a service at some time
point and how much time the service has spent to response.

\begin{figure}
  \centering
  \includegraphics[width=0.8\textwidth]{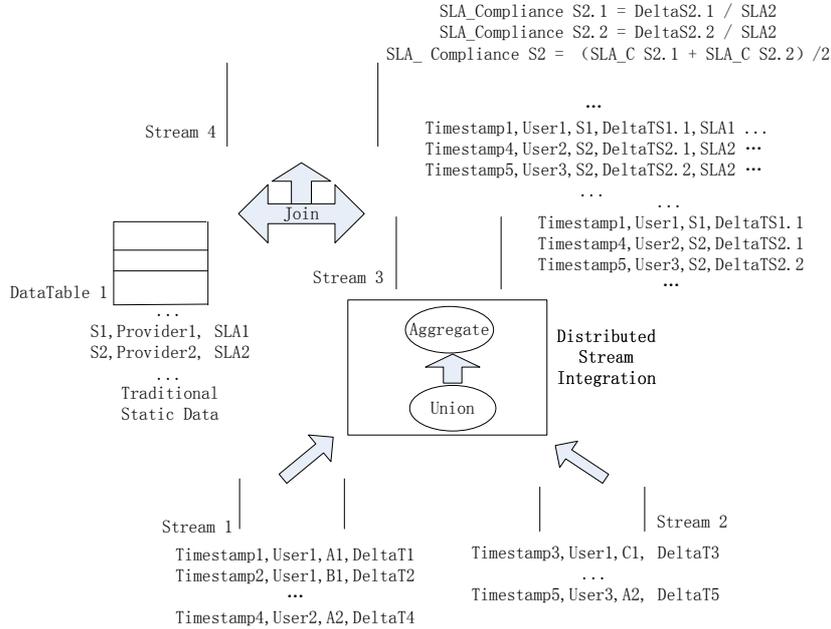}\\
  \caption{Case Study of Data Stream Integration for Recommendation}\label{Fig.15}
\end{figure}

Good services can be recommended to a new user by simply sorting SLA
compliances of the available services, while for old users, their
interest or preference may should be taken into consideration as
well. Given data streams of service invocations, to get the SLA
compliance of each service, we have to analyze the whole instance
which reflects the service quality and behavior in a service
process. The data tuples that record invocation information and
belong to the same instance are, as explained above, distributed
across multiple streams and may be disordered in invocation time.
Discussing how to integrate these data tuples distributed over
different data streams into one entire instance as accurately,
completely and effectively as possible is the key point of this
paper. As shown in figure \ref{Fig.15}, service $S_1$ is invoked
once, service $S_2$ twice. The invocation tuples are distributed in
the two data streams $stream_1$ and $stream_2$. The problem is how
to integrate correlated invocation tuples in $stream_1$ and
$stream_2$ to the service instance in $stream_3$.

Provided that we can address the issues well and obtain $stream_3$,
we are able to calculate the SLA compliance via joining and
comparing the conformity degree between the runtime measurement and
promised quality as shown in figure \ref{Fig.15}. Having the SLA
compliance of each book search service, it is feasible to recommend
befitting services to users through sorting relies on some rules and
is performed at some frequency.

\subsection{Invocation Association Accuracy} \label{subsec.association}

Associating relevant invocations distrubuted in different streams is
the fundamental step of integration, which plays a significant role
for further calculation of SLA compliance or service intelligence
model like recommendation and management. At the same time, it's not
such an easy task. First of all, because of the composite structure
and multi-copy properties of a service, the invocations of an
instance may be scattered in different streams. Second, if a
composite service is invoked several times, there will be multiple
identical invocations about each sub-service. How can we associate
separate but relevant invocations together into one instance
accurately? How can we distinguish the invocations of the same
service?

\begin{figure}
  \centering
  \subfigure[]{
    \label{Fig.04:a}
    \includegraphics[width=0.40\textwidth, scale=0.5]{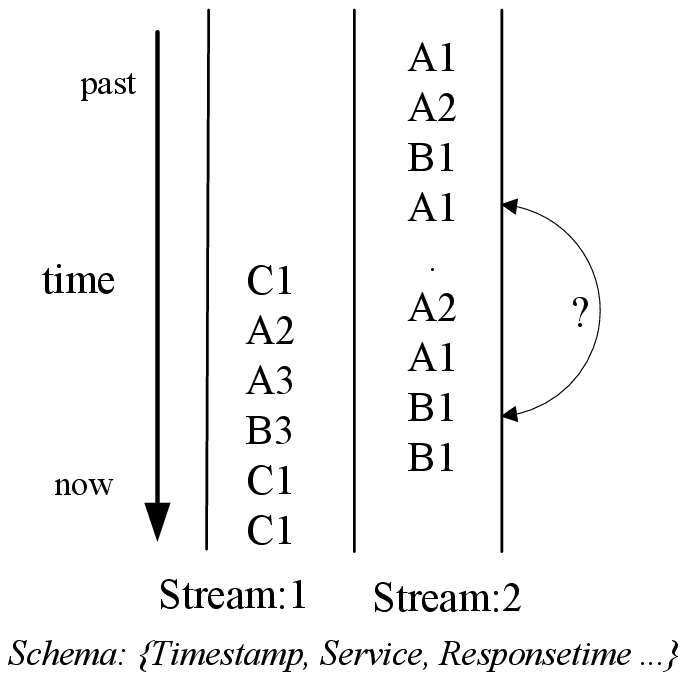}
    }
  \quad
  \subfigure[]{
    \label{Fig.04:b}
    \includegraphics[width=0.40\textwidth, scale=0.5]{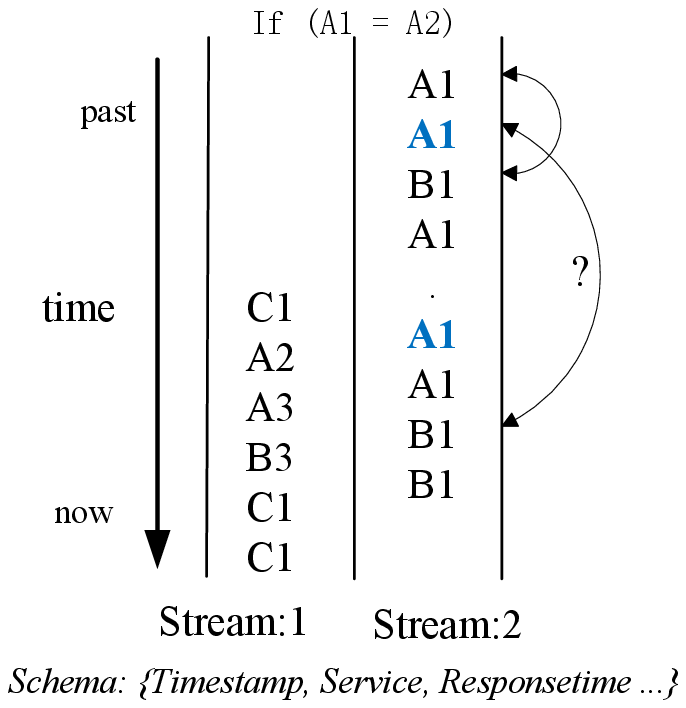}
    }
  \caption{Distributed stream illustrations}
  \label{Fig.04}
\end{figure}

Take the case in section \ref{Sec.Case} for example, we may get
event streams as shown in figure \ref{Fig.04:a} as time goes on. In
the monitoring streams, the requested service name - say $A_1$,
$B_1$, $C_1$ - is used to represent an invocation tuple. A tuple
usually includes a set of attributes such as timestamp, service
name, response time and so on. Intuitively, an tuple implies a user
invocation and records the service behavior as well as state
information which will be useful for further service intelligence
computation. As figure \ref{Fig.04:a} shows, $S_1$ and $S_2$ are
invoked for three times and $S_3$ is requested only once.We may
infer that the first appearance of $A_1, B_1$ in stream 2 and the
first $C_1$ in stream 1 belongs to an instance. But as for the
subsequent appearance of $A_1, B_1, C_1$, it is won¡¯t be easy. Are
the $n-$th $A_1$ and $n-$th $B_1$ in stream 2 pertaining to the same
instance? If $A_1$ and $A_2$ are the same service (this implies that
atomic service $A_1$ acts as a service component for several
services) as depicted in figure \ref{Fig.04:b}, how can we associate
the invocations exactly? Such questions are worth exploring.

\subsection{Information Integration}
\subsubsection{Integration Completeness}
 Supposing that we can address the association problem
proposed in section \ref{subsec.association} well, we may want to
gather as much invocations of an instance as possible for further
modeling. The instance summary information reflects an independent
service's behavior and state during a service process and can be
used as an important measurement for quality of service and SLA
compliance. Consequently, it is of great significance to integrate
the relevant invocations as completely as possible. However,
distributed data streams own intrinsic issues - dispersity and
asynchronism. How can we overcome such a tough nut?

\begin{figure}
\begin{center}
  \includegraphics[width=0.6\textwidth,scale=0.5]{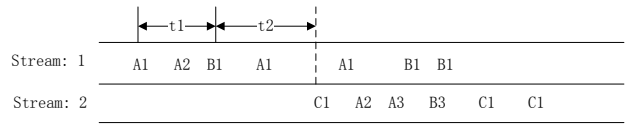}\\
  \caption{Interval between relevant invocations in distributed streams}\label{Fig.05}
\end{center}
\end{figure}

Dispersity indicates invocations pertain to the same instance are
scattered over multiple streams. Asynchronism means the relevant parts in
different streams don't arrive simultaneously. As figure
\ref{Fig.05} shows, the isolation of the first $A_1$ and $B_1$ by
$A_2$ in stream 1 illustrates the dispersity while the interval
between $A_1$ of stream 1 and $C_1$ in stream 2 indicates the
inter-stream asynchronism. Here our target is to integrate the
related invocations - say $A_1, B_1, C_1$ - into a corresponding
instance as completely as possible.

Mostly, the challenge arise from the instantaneity of streaming
data. Once a tuple is processed by a streaming operator, it will be
sent to the destination or discarded without any local copy. In
normal stateless data stream, a tuple has no subordinate or
membership relation with other tuples. Even those tuples with the
same key are only description for the same object's different
states. As for our case, the related invocation tuples belonging to
the same instance are scattered in multiple streams now. Usually, a
stream may include several invocation tuples of the same instance.
As a result, a tuple in the stream has something to do with some
other tuples before or after itself. That's to say, our data streams
are stateful. The state property of streaming data make it harder to
integrate completely. By the way, the underlying streams of
real-time information or functionality service could be either
stateful or stateless. In other words, problems included in this
real-time scenario are no more complex than those in multi-spot
service monitoring scenario.

\begin{figure}
\begin{center}
  \includegraphics[width=0.6\textwidth,scale=0.5]{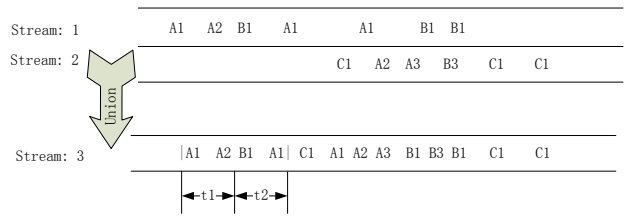}\\
  \caption{Union of distributed streams}\label{Fig.06}
\end{center}
\end{figure}

Data is very likely to get lost due to the dispersity and
asynchronism. In order to gather relevant tuples in stateful streams
as completely as possible, we may need an appropriate buffer
mechanism. Stream operators that rely on the relation between tuples
and employ sliding window to perform data processing
\cite{elsarticle-num:babcock_models_2002,
elsarticle-num:golab_issues_2003} like JOIN, AGGREGATE will be our
first choice. However, JOIN operator doesn't fit right here because
of the inter stream asynchronism. In fact, as shown in figure
\ref{Fig.06}, we first \textbf{\emph{UNION}} the distributed streams
into one rich stream and then \textbf{\emph{AGGREGATE}} relevant
invocation tuples to instances. The AGGREGATE operator will be the
emphasis of our study.

In round figures, the completeness has something to do with the
window mechanism and parameters. Take figure \ref{Fig.06} for
example, if the window size is set to be 4, then the tuple $C_1$
will be lost when considering the first invocation of $S_1$.
Increasing the window size may make the first instance of $S_1$
complete but the subsequent service instances remain incomplete. How
to obtain appropriate parameters? Is enlarging window size
reasonable for improving integration completeness? Is the sliding
window mechanism suitable for distributed stream data intehration?
These are the issues we are going to address in this paper.

\subsubsection{Performance Evaluation}

The major concern of performance evaluation during integration is
the performance measurement and resource consumption of each stream
operator. Stream operators are going to make up an operator network
of certain structure. The behavior analysis of single operator is
the cornerstone of stream processing performance analysis and
strategy decision making.

For example, if we need to merge multiple data streams, we may
wonder how long will the UNION operator take to forward a tuple?
Will the operator be blocked for the time-consuming operation and
hence result in tuple losing? As for the stateful AGGREGATE
operator, it quite a different story. It usually open a sliding
window to buffer incoming tuples and process them in a batch mode.
In this way, how long will it take to process a window? Can we have
a rough estimate about the memory consumption of the operator
according to the stream rate? Usually a new window will be be open
if the original one is filled and there may be multiple windows in
the system at the same time, so will the AGGREGATE operator run out
of the system memory?

To evaluate whether a integration strategy is good or bad, in
addition to the accuracy and completeness of the final result, we
also need to consider its performance, resource consumption in order
to give a more comprehensive evaluation. In this paper we will use
the queuing theory as the theoretical basis to model the data stream
operators, and present detailed analysis.

\subsection{Design of Experiment}
In practice, a web page is quite similar to a composite service.
Intuitively, the HTML file is equivalent to the head service while
the objects on a page are akin to subordinate services. A request
for an object on a page will fill the "\textbf{\emph{referrer}}"
field in the request header to declare which page the current object
belongs to. When a page is requested for several times, multiple
identical instance of the page will be generated. In addition, we
also get to know the objects of web pages on Amazon website are
mainly located in two domains. The static files such as HTML, JS and
CSS are stored in a domain while images with kinds of format, like
PNG, JPEG and GIF, are located in another region. The distributed
storage goes well with the distributed data streams scenario.

Therefore in our experiment, we use the Amazon access trace obtained
through HtmlUnit (an open source java web browser)
\cite{elsarticle-num:htmlunit} as our data set. The data we
collected includes invocation sequence, request time, requested
object, the referred web page etc. To make the data more convincing,
we consulted the interval time distribution of the English wiki
trace \cite{elsarticle-num:Wiki_Trace, elsarticle-num:wikimedia}
between 2008-1-1 22:50:22 and 2008-1-1 23:50:22. We first obtained
the interval request time distribution from wiki and then generate
several homogenous random sequences that have identical
distribution. We also find that there are 270,000 unique pages among
the 440,000 page instances in the specified trace. Taking for example
the multiplicity of enwiki's page instances, we finally acquired
13,997 instances from the original 10,000 unique pages in the Amazon
trace.

\begin{figure}
  \begin{center}
  \includegraphics[width=0.6\textwidth,scale=0.5]{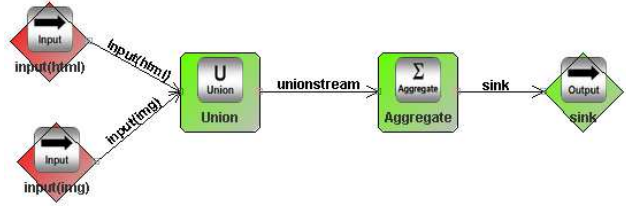}\\
  \caption{Basic stream operator networks}\label{Fig.07}
  \end{center}
\end{figure}

Based on the distributed storage environment mentioned above, we
assume that for each storage region, a stream conforming to the same
schema will be generated. During the data processing, we can first
UNION the two streams into a rich stream and then AGGREGATE the
objects that belong to the same page to obtain page instance level
summary information. Finally, the instance-level tuples are sent to
the sink node. The stream processing strategy is shown in figure
\ref{Fig.07} and the computation node configuration is shown in
table \ref{Tab.02}. Note that the operators can be deployed on the
same node or separately.

\begin{table}
  \centering
  \caption{Experiment environment}\label{Tab.02}
  \small
  \begin{tabular}{cc}
    \hline
    CPU & Intel Pentium Dual CPU E2160 1.8GHz \\
    Memory & 1979MB \\
    OS & Linux Ubuntu 9.04 \\
    Tool & Borealis Software, Matlab7.0, MySQL 5.1 \\
    \hline
  \end{tabular}
\end{table}

\section{Association Strategy}

In this section, we will first give a brief introduction to
information types that could contribute to identifying invocations
and then propose potential associate strategies systematically.

\subsection{Information Type}
\begin{enumerate}
  \item \textbf{\emph{Service Information}}

  Service information contains the service identifier and
service structure. A service identifier is a unique symbol which distinguishes
different services and can be represented in the form of service
names, digital ID, URL, etc. Service structure tells the sub
services of the current one or which service the current one belongs
to (it is possible that an atomic service can belong to multiple
services). In practice, service structure information can be
distributed to clients through responses and then will be included
in the subsequent requests. For example, the HTTP request header
usually contains a "\textbf{\emph{referrer}}" entry, which is the
URL of web page that the current requested object belongs to.

  \item\textbf{\emph{Client Information}}

Client information can be the identifier of client agent (e.g. IP
address) or user-specific information such as cookie. Usually it can be obtained in the
request header and do good to distinguishing different users but is
useless for seperating multi-instances of a service generated by the same
person.

  \item\textbf{\emph{Instance Status Information}}

An instance will be produced when an independent service is invoked
and it usually contains some dynamic status information like
invocation sequence number (ID) and timestamp. The instance ID can
distinguish instances perfectly if it is unique for each instance
but also leads to great overhead for maintaining uniqueness, and in
fact it is not practical in distributed environment. In this paper,
we mainly employ timestamp as the status information to help
differentiate instances. For each primary invocation, a timestamp
will be generated and encapsulated into the response; as for
subordinate invocations, the information are contained in the
request headers. Keep in mind that in this paper, we use the
timestamp of head service to represent that of the entire instance
for convenience.
\end{enumerate}

\subsection{Association Strategies}
In the following subsections, we will illustrate different associate
strategies by employing the above three kinds of information. As
shown in Figure \ref{Fig.07}, all the strategies involves UNION and
AGGREGATE. UNION puts multiple streams together into one stream and
AGGREGATE groups the invocation records into instances according
specific key attributes. The data stream processing goes smoothly
provided that the streaming engine has an infinite buffer.

\subsubsection{Service Structure Information + Instance Status Information (Timestamp)}
As shown in Table \ref{Tab.04}, "Head service" and "Instance
Timestamp" are used as the association key. Keys (A, 2010-12-21
17:11:29) and (A, 2010-12-21 17:11:37) stand for different instances
of composite service A. If the load is stable, this strategy can
distinguish instances of the same or different services. But if many
a user invoked the same service within a single time unit, the
instances of a service will have the same timestamp which leads to
the failure to distinguish.

\begin{table}
  \centering
  \caption{Stream structures of "Head + Timestamp"}\label{Tab.04}
    \small
    \begin{tabular}{cccc}
    \toprule
    Requested Service & Head Service & Instance Timestamp & Other Information \\
    \midrule
    B & A & 2011-2-25 17:11:29 & \ldots \\
    B & A & 2011-2-25 17:11:29 & \ldots \\
    B & A & 2011-2-25 17:11:29 & \ldots \\
    \bottomrule
  \end{tabular}
\end{table}

\subsubsection{Service Structure Information + Client Information}
In table \ref{Tab.05}, "Head service" and "Client information" are
used as the identify keys. Tuples (A, 192.169.10.28) and (A,
192.168.10.111) represents different instances of service A. This
scheme can be used to distinguish different users for services.
However, the strategy cannot discriminate multi-instances of the
same service generated by the same user during a specified interval.

\begin{table}
  \centering
  \caption{Stream structures of "Head + Client"}\label{Tab.05}
  \small
  \begin{tabular}{cccc}
    \toprule
    Requested Service & Head Service & Client Information & Other Information \\
    \midrule
    B & A & 192.168.10.28 & \ldots \\
    B & A & 192.168.10.111 & \ldots \\
    B & A & 192.168.10.28 & \ldots \\
    \bottomrule
  \end{tabular}
\end{table}

\subsubsection{Service Structure + Instance Status Information (Timestamp) + Client Information}
This is a re-enforced version of "Service structure + Timestamp"
strategy and its stream structure is shown in table \ref{Tab.06}.
The only difference is that the number of associated attributes
increases from two to three, which may increase the association cost for each
tuple. With the additional "Client information", it can
distinguish multiple instances of the same service invoked by
different clients within a single time unit and the error is
extremely small. An exception is that a strange client invoked the
same service within a single time unit for several times.

\begin{table}
  \centering
  \caption{Stream structure of "Head + Timestamp + Client" strategy}\label{Tab.06}
  \small
  \begin{tabular}{cccc}
    \toprule
    Requested Service & Head Service & Timestamp & Client Information \\
    \midrule
    B & A & 2011-2-25 17:11:29 & 192.168.10.28 \\
    B & A & 2011-2-25 17:11:37 & 192.168.10.111 \\
    B & A & 2011-2-25 17:11:29 & 192.168.10.28 \\
    \bottomrule
  \end{tabular}
\end{table}

\subsection{Brief Summary}

\newcommand{\tabincell}[2]{\begin{tabular}{@{}#1@{}}#2\end{tabular}}
\begin{table}[!htbp]
  \centering
  \caption{Brief Summary\tnote{1}}\label{Tab.03}
  \small
  \begin{threeparttable}
  \begin{tabular*}{1.0\textwidth}{l}
    \toprule
    \textbf{"Service Structure + Client Information"} \\
    \tabincell{l}{(\checkmark): If users access the same service in sequential mode(there is no limitation for different services)} \\
    (\dag): It can't tell the instances of the same service that
generated by concurrent access of the same user \\
    \hline
    \textbf{"Service Structure + Timestamp"} \\
    (\checkmark): Under normal work load, it has good discriminatory capacity \\
    \tabincell{l}{(\dag): If a service is invoked for several times in a time unit or has multi-copies running on a platform,\\ \qquad the
multi-instances of the same service may have the same timestamp and
can't be
distinguished} \\
    \hline
    \textbf{"Structure + Timestamp + Client Information"} \\
    (\checkmark): It is the most powerful approach and can cover the shortcomings of
the above two methods \\
    (\dag): It may  should pay cost for the performance. \\
    \bottomrule
  \end{tabular*}
  \begin{tablenotes}
  \item [1] {(\checkmark) $-$ applicable scenario; (\dag) $-$ non-applicable scenario}
  \end{tablenotes}
  \end{threeparttable}
\end{table}

\section{Integration Completeness}\label{Sec.Integration}
As shown in figure \ref{Fig.07}, our integration approach consists
of two stream operators: UNION and AGGREGATE. The stateless UNION
operator merges more than one identical streams into a rich one and
the stateful AGGREGATE employs window mechanism to perform relevant
computation. As for the window type, there are two typical classes
of sliding windows: counting and timing window. The counting window
accommodates a fixed number of tuples while the time window can hold
all the tuples in a specified interval. In this paper, our main
efforts are directed to reasonable window mechanism as well as its
parameters in AGGREGATE operator to maximize the information
integration completeness.

\begin{figure}
  \centering
  \includegraphics[width=0.4\textwidth,scale=0.5]{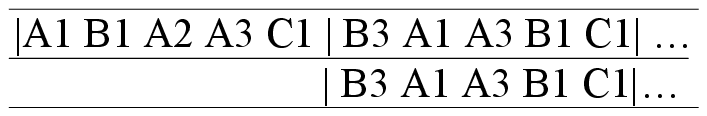}\\
  \caption{Invocation event stream}\label{Fig.08}
\end{figure}

In general, the normal sliding window performs pool in our stateful
stream scenario. Supposing there is a set of services $\{S_1, S_2
\dots S_k\}$ and each service has its own degree. For instance,
service $S_1$ includes sub-services $A_1$, $B_1$, and $C_1$; $S_2$
only contains $A_2$; $S_3$ includes $A_3$ and $B_3$. At some time,
the snapshot of a stream is shown in figure \ref{Fig.08}.

Assuming that both the window size and the advance step has the
value of 5 we can get the partition in figure \ref{Fig.08}. We can
get two intact instances - ($A_1$, $B_1$, $C_1$) and ($A_2$) - from
the first window and A3 is lost. In the second one we only get one
intact instance - ($A_1$, $B_1$, and $C_1$). In total, three out of
ten invocation tuples get lost in this partition.

Some may argue that the reasonable advance step can do good to
improving the completeness of integration. We first should be aware
that the advance step is fixed and can't be changed dynamically. The
adjusting of advance step may work well under some conditions but
still bad in other scenarios. Because we can't know well about the
arrival pattern of stream in advance and change the step adaptively,
the performance of sliding window depends heavily on luck. In
addition, increasing the window size may reduce the information loss
rate at some degree but couldn't solve the problem fundamentally.

\subsection{Small Window Array Mechanism AGGREGATE}
Aiming at improving the integration completeness in stateful
streams, we introduce a small window array mechanism that fit well
in our scenario. It works as follows:

The operator box maintains a map of small windows at runtime. A
$<$key, value$>$ pair represents the value of association attributes
and the corresponding small window in which all the tuples have the
same key. The operator box extracts each tuple's key at its arrival.
If there is no such key in the map, the operator box will open up a
new small window with fixed size, insert the tuple into the window
and finally put the new key and window pair into the map. If such
key exits, the operator box just insert the invocation tuple into
the relevant window. If a small window is full, it will be closed
and calculated automatically. Timeout is also supported to close a
window which stays in the system for an unreasonablly long time.
When a window is constructed, it will have its first tuple
immediately. This is the time point that the clock start to timing.

\begin{figure}
  \centering
  \includegraphics[width=0.40\textwidth,scale=0.50]{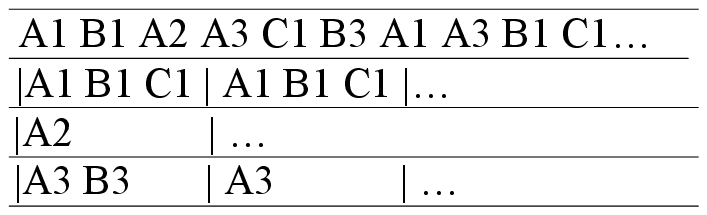}\\
  \caption{Small window array illustrations}\label{Fig.09}
\end{figure}

A case in point of small window array is shown in figure
\ref{Fig.09}. The first row represents the tuple sequence, and the
other rows stand for small windows for service instances. The
invocation tuples in a small window belongs to a service (page)
instance. We refer to a small window as an instance for convenience.
Here the biggest service degree is $|$S1$|$ = 3. If we set the
window size to 3, the information loss rate will be minimized to
zero. In practice, maybe only a small fraction of services have
large degree number and hence we can't use the biggest degree as the
small window size by default because of the potential waste of large
storage space.

BOREALIS maintains a window list for each key. We make sure
each list has no more than one window by setting window size equal
to advance step in the paper. Window parameters have great influence
on the outcome of information integration. In order to improve the
integration completeness and minimize the loss rate, we have to pay
great attention on setting rational window parameters values.

\subsection{Parameter Estimation: Counting Window Size}
We are going to obtain the counting window size in this way: first
of all, try to get the frequency distribution of service degree
through statistical analysis; then employ the
Expectation-Maximization \cite{elsarticle-num:PH_EM} method to
obtain the probability density function \textbf{\emph{g(n)}} and
cumulative distribution function \textbf{\emph{G(n)}}
\cite{elsarticle-num:EM_MAP}. The symbol \textbf{\emph{n}} means the
service degree, \textbf{\emph{g(n)}} indicates the percentage of
services that has a degree of $n$, \textbf{\emph{G(n)}} stands for
the proportion of the services whose degree is less than or equal to
$n$. Given the precise integration completeness $\alpha$, we can
acquire the appropriate $n$, which will be used as the window size,
by solving the equation F = $\alpha$.

It should be noted \textbf{\emph{F}} is the \textbf{\emph{precise
completeness}}. In practice, we can use \textbf{\emph{approximate
completeness}} which is based on level $\gamma$ to measure the
goodness of a parameter setting. Supposing the counting window size
is set to be \textbf{\emph{n}} and some service have a degree of
\textbf{\emph{N}}, the number of tuples collected in a small window
is $K(1 \leqslant K \leqslant n)$. When the window is closed, if $K
/ N \geqslant \gamma$ , we hold the opinion that this instance is
approximately integrated based on level $\gamma$ . In general, the
approximate completeness is higher than the precise one. They equal
to each other on condition that $\gamma = 1$. In fact, $1 - \alpha$
is considered to be the up limits of information loss rate if we use
approximate completeness to measure the outcome.

\begin{figure}
  \centering
  \subfigure[PDF of service degree]{
    \label{Fig.10:a}
    \includegraphics[width=0.34\textwidth, scale=0.5, angle=270]{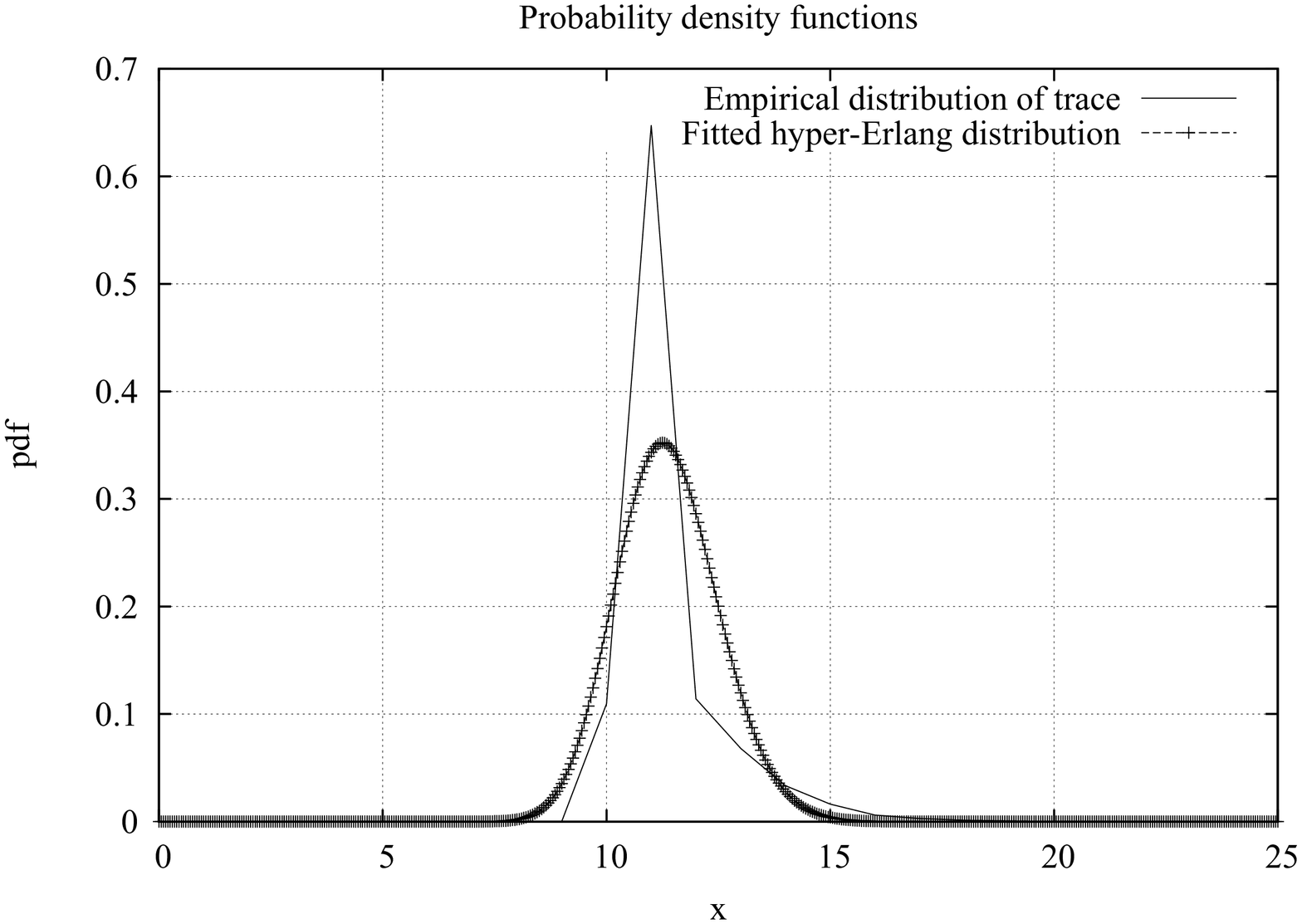}
    }
  \subfigure[CDF of service degree]{
    \label{Fig.10:b}
    \includegraphics[width=0.34\textwidth, scale=0.5, angle=270]{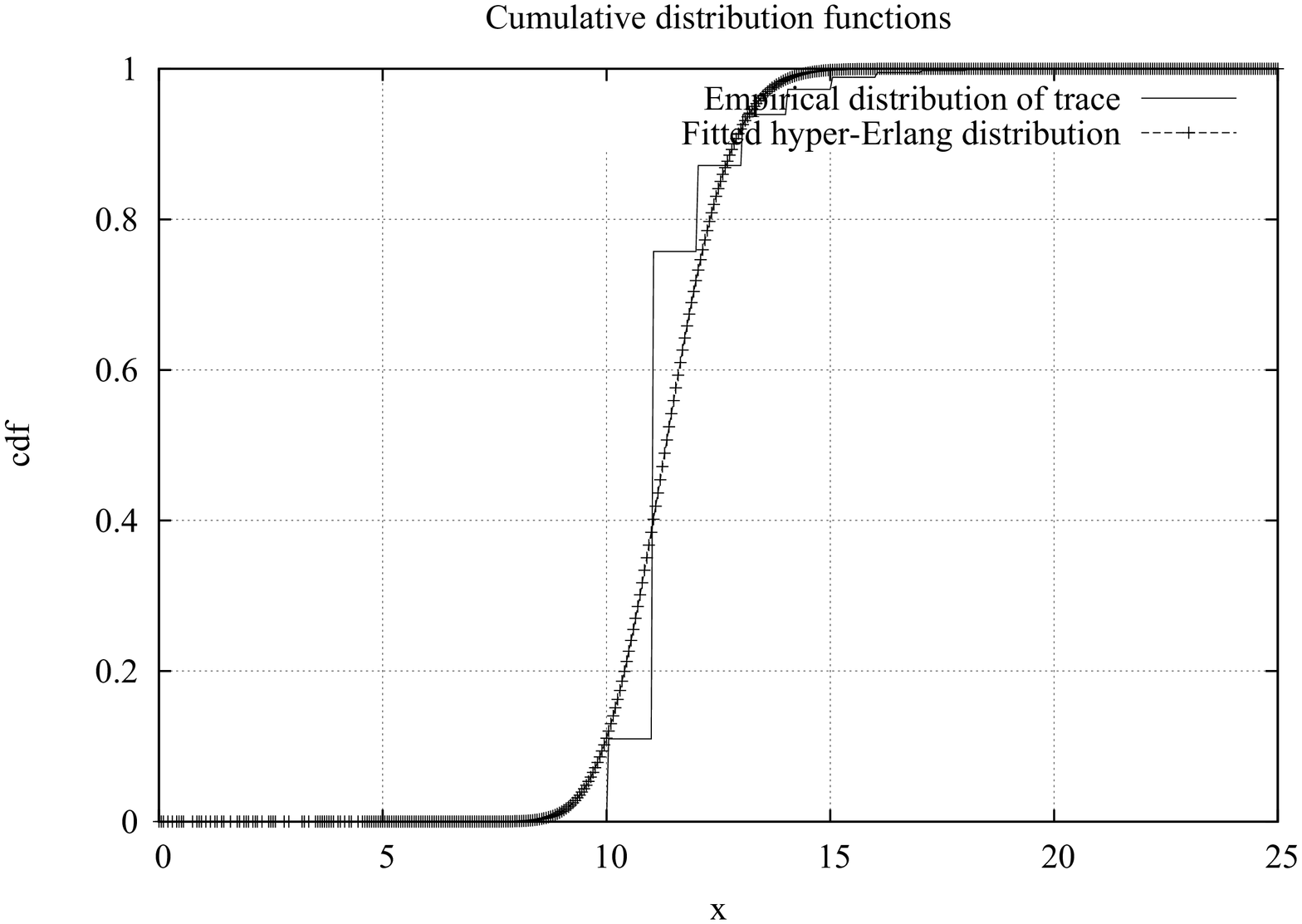}
    }
  \caption{PDF and CDF of service degree}
  \label{Fig.10}
\end{figure}

In our experiment, the service (page) degree is a single-branch
Erlang distribution with $\alpha = 1, E(\lambda,K) = E(8.7963, 100),
m = 1$. It can be expressed as formula (\ref{Formula.03}) and
depicted as figure \ref{Fig.10}. In figure \ref{Fig.10}, the light
gray line is the practical distribution while the dark gray
represents the fitted one. From it we can see that the majority of
service degree lie in $[10, 15]$ and the services whose degree are
less than or equal to 15 account for nearly $99\%$.

\begin{equation}\label{Formula.03}
\small
f_{degree}=
\frac{8.7963^{100}x^{99}e^{-8.7963x}}{\prod\limits_{i=1}^{99}i} =
2.8840^{-62}x^{99}e^{-8.7963x} \quad F_{degree}= 1 -
{\sum_{n=0}^{99}
\frac{e^{-8.7963x}{8.7963x}^{n}}{\prod\limits_{j=1}^{n}j}}
\end{equation}

Given integration completeness $\alpha$, the window size $n$ can be
obtained by solving $F_{degree} = \alpha$. In practice, there is a
little trouble when trying to obtain the analytic solution through
the \textbf{\emph{newton iteration}} method because of the
accumulation of multiple branches. However, if we can realize that
the window size $n$ is a natural number and definitely has a value
lies in $[0, max(degree)]$, things will be much simple. In fact, we
at last hire the \textbf{\emph{trial-and-error}} method to discover
the closest window size according to the given $\alpha$. For
example, supposing $\alpha = 0.90$, we can obtain the best suitable
window size $n = 13$. When $x_1 = 12, F_1 = 0.7186; x_2 = 13, F_2 =
0.9200; x_3 = 14, F_3 = 0.9857$. So $n = x_2 = 13$ is the best
choice for $\alpha = 0.90$.

\subsection{Parameter Estimation: Timing Window Size \& Timeout}
\begin{figure}
  \centering
  \subfigure[PDF of service response time]{
    \label{Fig.11:a}
    \includegraphics[width=0.34\textwidth, scale=0.5, angle=270]{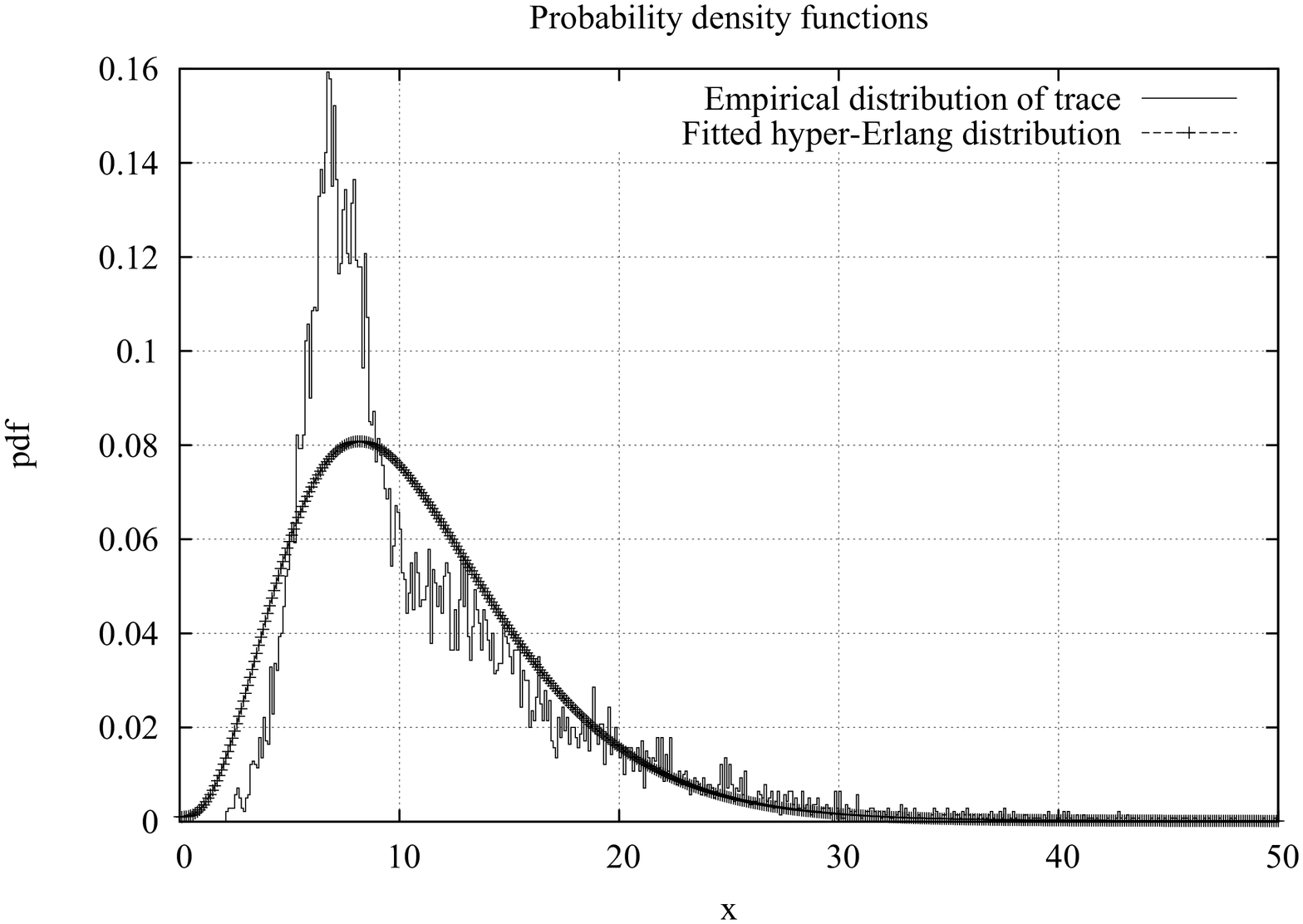}
    }
  \subfigure[CDF of service response time]{
    \label{Fig.11:b}
    \includegraphics[width=0.34\textwidth, scale=0.5, angle=270]{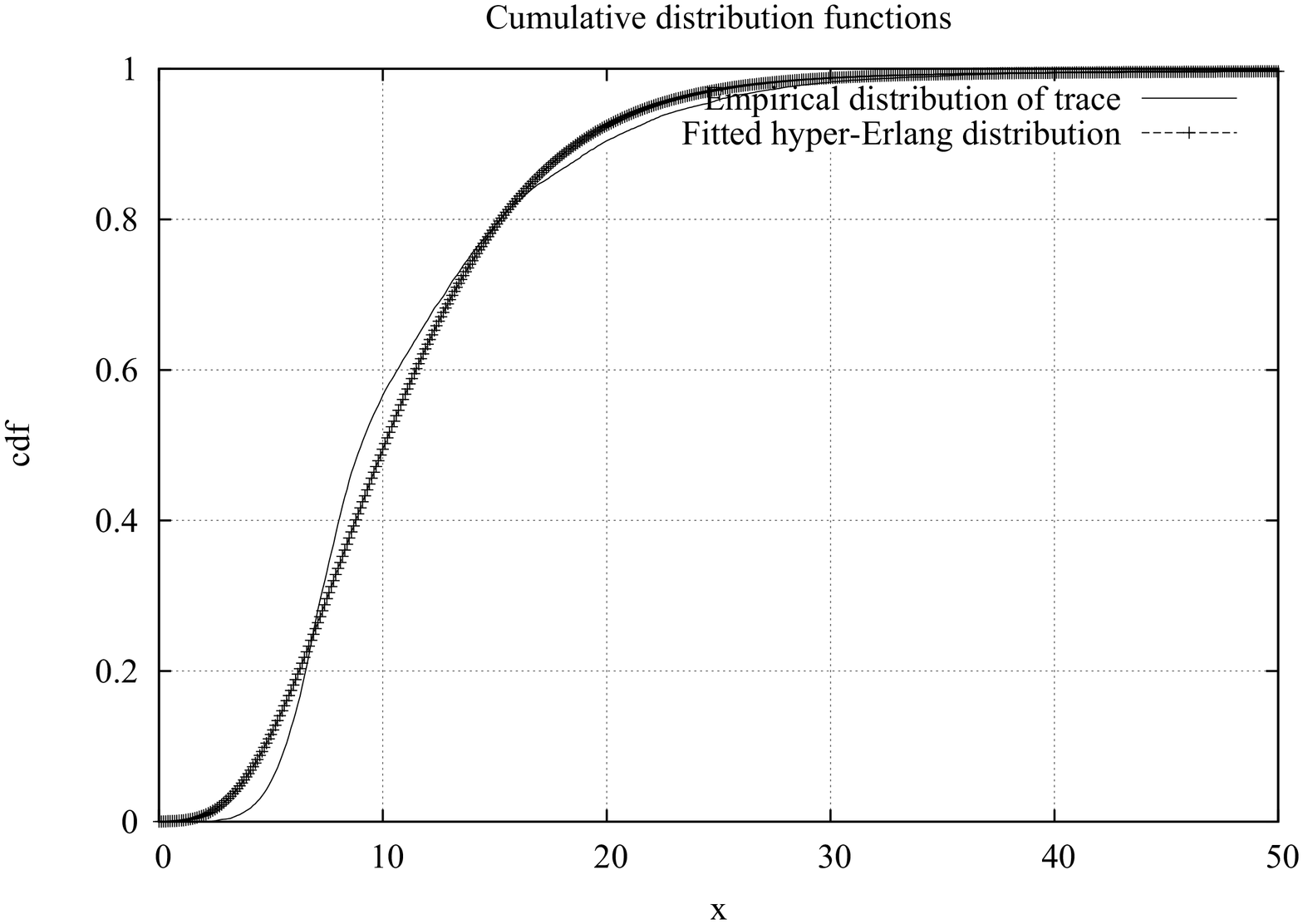}
    }
  \caption{PDF and CDF of service response time}
  \label{Fig.11}
\end{figure}

Similarly, the response time of services (pages) can be used to
guide the setting of time window and timeout. In our experiment, we
obtained a two-branch Hyper-Erlang distribution as shown in formula
(\ref{Formula.04}) and figure \ref{Fig.11}. The average response
time is $11.2513(s)$ and nearly $90\%$ of the services' response
time lasts no more than $20(s)$ (see figure \ref{Fig.11}).

\begin{footnotesize}
\begin{equation}\label{Formula.04}
f_{time}= 0.0009979e^{-0.0404x} + 0.0029x^{3}e^{-0.3666x}
 \quad F_{time}= 1 - (0.0247e^{-0.0404x} + 0.9753\sum_{n=0}^{3}\frac{{e^{-0.3666x}{0.3666x}^{n}}}{\prod_{j=1}^{n}})
\end{equation}
\end{footnotesize}

Similar with the analysis of counting window size, given precise
completeness $\alpha$, we can acquire the reasonable time window
size as well as timeout through $F_{time} = \beta$. For example,
given timeout rate $\beta = 0.05$, we can obtain a sound timeout
value $t = 22 (s)$. It means that around $5\%$ of the windows will
be closed forcibly if it stays over $22(s)$. In an ideal word,
$95\%$ of the service instances will be covered.

\subsection{Experiment Verification}
\subsubsection{Comparison of Window Mechanisms}
The association attributes are identical for both sliding window and
small window array mechanisms, including \textbf{\emph{service
structure}}, \textbf{\emph{instance status (timestamp)}} and
\textbf{\emph{client information}}. For small window array, given
precise completeness $\alpha = 0.90$ and timeout rate $\beta =
0.05$, we get window size $n = 13$ and timeout $t = 22 (s)$. The
normal sliding window size is taken as 8000, 16000 and 32000
(advance step = window size). The results are shown in table
\ref{Tab.07}.

\begin{table}[th]
  \centering
  \begin{threeparttable}
  \caption{Comparison between sliding window and small window array (SWA) mechanisms}\label{Tab.07}
  \footnotesize
  \begin{tabular}{cccccc}
    \toprule
    Indicators & Small Window Array & \multicolumn{3}{c}{Sliding Window} \\
    \midrule
    Window Parameters \tnote{a} & 13/13/22 & 8000/8000 & 16000/16000 & 32000/32000 \\
    Integration Completeness($\gamma=1$) \tnote{b} & 80.48\% & 13.15\% & 43.44\% & 72.11\%\\
    Integration Completeness($\gamma=0.85$) & 90.30\% & 15.17\% & 47.63\% & 74.73\% \\
    Integration Completeness($\gamma=0.75$) & 91.66\% & 16.99\% & 45.87\% & 73.76\% \\
    Integration Completeness($0<\gamma\leqslant1$) \tnote{c} & 94.06\% & 40.16\% & 63.15\% & 76.27\% \\
    \bottomrule
  \end{tabular}
  \begin{tablenotes}
  \item [a]{The window parameters means window size/ advance step/ timeout(if necessary)}
  \item [b] {If $\gamma=1$, the instance is integrated completely}
  \item [c] {If $0<\gamma<1$, the instances is integrated partially}
  \end{tablenotes}
  \end{threeparttable}
\end{table}

Figure \ref{Tab.07} shows that the small window array mechanism performs
much better. Its precise completeness is $80.4815\%$ when
$\gamma=1$, approximate completeness based on $\gamma=0.85$  and
0.75 reaches $90.2979\%$ and $91.6553\%$ respectively. In addition,
$94.0649\%$ of the invocations have been integrated (once an
invocation is gathered into an instance, the completeness of that
instance will be bigger than 0).

When it comes to the normal sliding window mechanism, the integrate
completeness increases with the enlargement of window size. Even
though the sliding window mechanism has similar completeness when
the window size reaches 32000, there are still several intrinsic
problems in it. First of all, the completeness here is obtained
after a serial of experiments. In fact, data stream can only be
handled once at each operator node and hence more than one attempt
for reasonable window parameters is impossible. Second, the sliding
window will consume more resources (please refer to table
\ref{Tab.12} for details). Another issue is that the window size is
nearly one-sixth of the total tuple records (174069) which make the
outcome unconvincing for in theory the completeness will reaches
$100\%$ if the window size is equal to the entire data set size. In
practice the window size is quite limited when comparing to the
infinite length of data streams.

\subsubsection{Quantitative Analysis of Association Strategies}
We proceed to analyze the association strategies based on the small
widow array approach with identical parameter settings $13/13/22$
here. From table\ref{Tab.08} we can see that the "Head" strategy
which only makes use of service structure information performs
relatively bad. The recall rate of service (page) instances is
91.3053\% and the accuracy is 85.1683\%. The "Head + IP" strategy
which relies on service structure and client information performs
pretty well in the ordinary case as we have expected. Provided that
users access the same service in a sequential order (there is no
limitation about different services), this approach can be
considered to be the first choice. Though the "Head + Timestamp +
IP" strategy owns the best effectiveness, it also consumes extra
computation resource and capacity. It is not necessary to pursue
little improvement at the high price of processing overhead.

\begin{table}[th]
  \centering
  \caption{Analysis for association strategies}\label{Tab.08}
  \begin{threeparttable}
  \footnotesize
  \begin{tabular}{ccccc}
  \toprule
  Strategy  &  Head  &  Head + Time & Head + IP & Head + IP + Timestamp\\
  \midrule
  Window Parameters  &  13/13/22 &   13/13/22 &   13/13/22 & 13/13/22 \\
  Integration Completeness($\gamma=1$)  & 70.38\% & 79.53\% & 80.44\% & 80.4815\% \\
  Integration Completeness($\gamma=0.85$)  &  78.30\% & 89.14\% & 90.13\% & 90.2979\%\\
  Integration Completeness($\gamma=0.75$) &  79.41\% & 90.48\% & 91.49\% & 91.6553\%\\
  Integration Completeness($0<\gamma\leqslant1$) &  81.14\% & 92.79\% & 93.98\% & 94.0649\% \\
  Recall\tnote{a} & 91.31\% & 99.31\% & 99.99\% & 100\%\\
  Correct rate\tnote{b} & 85.17\% & 98.63\%  & 99.98\% & 100\%\\
  \bottomrule
  \end{tabular}
  \begin{tablenotes}
    \item [a] {Ratio of the integrated instances and the total instances}
    \item [b] {If an integrated instance contains more than or equal to one invocation that doesn't belong it, we think the instance is associated improperly}
  \end{tablenotes}
  \end{threeparttable}
\end{table}

\section{Performance Evaluation}\label{Sec.Queuing}
In this section, we will employ queuing theory to carry on behavior
analysis for typical stream operators involved in the processing
strategy. The behavior or performance analysis has considerable
guiding significance for stream processing and operator network
optimization. For instance, queue and waiting queue length can be
used to predict the resource consumption; the probability that a
coming tuple will be rejected by the system is helpful for load
shielding.

\subsection{Obtaining Prior Parameters of Queuing Model}
In order to build a rational queuing model, insightful understanding
about the three major components - input process, service time and
service discipline - comes first. The input interval and service
time distribution should be obtained in anticipation during a
modeling process.

We obtained a practical web page request distribution from wiki
trace. It turns out to be a 2-stage PH distribution through
employing the distribution fitting approach
\cite{elsarticle-num:osogami_analysis_2005} and is expressed as
formula (\ref{Formula.05}):

\begin{equation}\label{Formula.05}
    \alpha = [1, 0] \qquad T = \left(
                                 \begin{array}{cc}
                                   -0.1452 & -0.0329 \\
                                   0 & -0.1191 \\
                                 \end{array}
                               \right)
\end{equation}

We try to generate several homogenous random sequences that have
identical distribution with wiki trace. At first we replay the
sample data into the stream system in the light of one of the
homogenous sequences to gain the service time distribution of a
specific operator. Once the input tuple arriving interval and
service time distribution are obtained, the model will be easy to
set up.

It should be noted that the original web page request distribution
has not taken the objects on pages into consideration. In other
words, the data stream is a primary invocation tuple stream. When
considering the subordinate invocation tuples, the stream will
become much denser and have different distribution coefficients. We
get another 2-stage PH Distribution after combining all the primary
and subordinate invocations. It can be expressed as formula
(\ref{Formula.06}):

\begin{equation}\label{Formula.06}
    \alpha = [1, 0] \qquad T = \left(
                                 \begin{array}{cc}
                                   -1.1215 & 0.0001 \\
                                   0 & -0.0021 \\
                                 \end{array}
                               \right)
\end{equation}

\subsection{Queuing Model for UNION Operator}
Union is stateless and serves as a router. It accepts several
$(\geqslant 1)$ streams that have a uniform schema and merges them
into a stream with identical schema according to tuples' arriving
order \cite{elsarticle-num:Borealis_Guide}. In our experiment, there
are two distributed streams. The input distribution of UNION is the
combination of the two branches and is the same as formula
(\ref{Formula.06}).

The service time conforms to a 4-stage PH distribution with average
service time $0.005364(ms)$. The service time distribution can be
recorded as formula (\ref{Formula.07}) :

\begin{equation}\label{Formula.07}
    \alpha = [1, 0, 0, 0] \qquad T = \left(
                                       \begin{array}{cccc}
                                         -378.3987 & 378.3987 & 0 & 0 \\
                                         0 & -378.3987& 378.3987 & 0 \\
                                         0 & 0 & -12669.0969 & -0.0000346 \\
                                         0 & 0 & 0 & -0.05120 \\
                                       \end{array}
                                     \right)
\end{equation}

\begin{figure}
  \centering
  \includegraphics[width=0.6\textwidth]{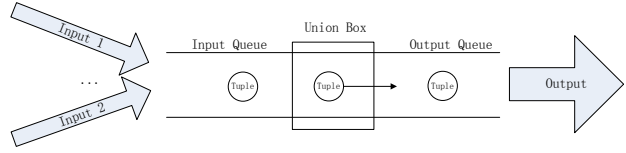}\\
  \caption{Architecture of Union operator}\label{Fig.12}
\end{figure}

To understand the working mechanism of the union operator box, it is
better to have a close look at the operator architecture. In figure
\ref{Fig.12}, the union accepts multiple identical streams and
inserts each incoming tuple of those streams into
\textbf{\emph{one}} input queue if there is enough space or drops it
otherwise. The first tuple in the waiting queue will be put forward
to output queue if the operator is idle or left alone for waiting
otherwise. In BOREALIS, the input queue length can be regulated
through changing the following parameters: AURORA\_PAGE and
AURORA\_PAGE\_SIZE. AURORA\_ PAGE means the number of memory pages
reserved for the queue and defaults to 4000; AURORA\_PAGE\_SIZE
indicates the size of each page and is 4096 bytes by default. Given
the size of each tuple, we can get the queue length.

In our experiment, AURORA\_PAGE and AURORA\_PAGE\_SIZE are set to be
10 and 1024 respectively. In addition, the size of a tuple is 135
bytes. The queue length can be calculated via expression
(\ref{Formula.08}). Getting rid of trivial storage overhead, we
obtained the final queue length at around 60.

\begin{equation}\label{Formula.08}
    \frac{AURORA\_PAGE \times AURORA\_PAGE\_SIZE}{Sizeof(Tuple)} = 10 \times \lfloor
    \frac{1024}{135}\rfloor = 70
\end{equation}

From the discussion above, we can build an appropriate queuing model
for the UNION operator. Of which the input process is a 2-stage PH
distribution, service time conforms to a 4-stage PH distribution,
the number of service desk is 1, the buffer size is 60, the service
discipline is FCFS and the number of input tuples is infinite. It
can be noted as $PH_{2}/PH_{4}/1/60/\infty/FCFS$.

Keep in mind that PH distribution is a kind of markovian arrival
process (MAP) and also a sort of general distribution. With the
background we can obtain our model via replacing $a$, $b$ with the
same value $1$ in  $MAP/G^{[a, b]}/1/N$ and get the leading
performance indicators by using the matrix-based approach
\cite{elsarticle-num:gupta_analysis_2001} in that model.

In table \ref{Tab.09}, the difference between the predicted average
process time 0.005388 (ms) and the observation 0.005370 (ms) is
quite small $(0.000018 ms)$ and the error rate is only 0.33\%. As
for the queue length, the predicted value says 0.005698 while
observation is 0. In addition, a coming tuple has to wait at a
probability of 0.5673\% (system is busy) or will be dropped at a
probability of 0.0027\%. We regard the model to be acceptable based
on the conformity between prediction and observation.

\begin{table}[th]
  \centering
  \caption{Behavior indicators of $PH_{2}/PH_{4}/1/60$ model}\label{Tab.09}
  \begin{threeparttable}
  \small
  \begin{tabular}{ccccc}
  \toprule
  \multicolumn{4}{l}{\textbf{Prediction:}}\\
  \midrule
  Queue Length (L) \tnote{a} &  0.005698  &  Waiting Queue Length (Lq) \tnote{b} & 0.000025\\
  Residence Time (W)\tnote{c} & 0.005388 &   Waiting Time (Wq) \tnote{d} & 0.000024 \\
  Pbusy \tnote{e} & 0.005673  &  Ploss \tnote{f} & 0.000027 \\
  \multicolumn{4}{l}{\textbf{Observation:}}\\
  Queue Length (L) &   0  &  Error &   $-$\\
  Residence Time (W) &   0.005370 &    Error &   0.33\% \\
  \bottomrule
  \end{tabular}
  \begin{tablenotes}
  \item [a] {Queue length, tuples in the operator box and waiting queue }
  \item [b] {Tuples in the waiting queue}
  \item [c] {Residence time, including the processing time and waiting time}
  \item [d] {Waiting time}
  \item [e] {The probability system is busy at a tuple arrival}
  \item [f] {The probability that a tuple is rejected at its arrival}
  \end{tablenotes}
  \end{threeparttable}
\end{table}

One thing to note is that we obtain the observation through chopping
off the head and feet of data streams which makes the observation
more closer to the practical infinite scenario. In addition, queuing
models for other stateless operators, like MAP, FILTER can be built
in a similar way.

\subsection{Queuing Model for Sliding Window AGGREGATE Operator}
The normal sliding window aggregate operator buffers a certain
number of tuples from the input stream and partitions them into
different groups according to their keys. For each group, the
operator will perform specified operation(s) on corresponding
attributes and produce one tuple
\cite{elsarticle-num:Borealis_Guide}. When defining an aggregate
operator, important parameters like window type, window size,
timeout, sorting attribute and aggregate function(s) should be
specified.

In our experiment, the input process of aggregate operator is the
same as UNION (refer to expression (\ref{Formula.06})). The normal
aggregate operator works in a batch mode. Assuming the waiting queue
length to be $N$, the operator will keep waiting until the number of
tuples in the queue exceeds $a$, and serve at most $b$ tuples each
time. In general, $a$ and $b$ are assigned with the same value -
window size $K$ and which means the operator will perform a
calculation each $K$ tuples.

\begin{figure}
  \centering
  \includegraphics[width=0.7\textwidth, scale=0.5]{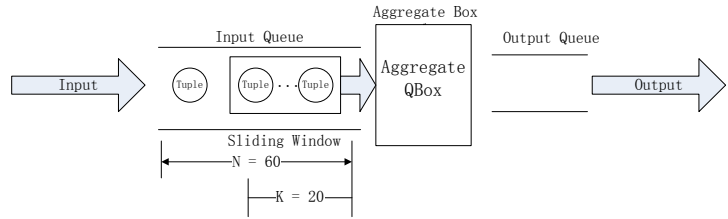}\\
  \caption{Architecture of Aggregate operator box}\label{Fig.13}
\end{figure}

The design architecture of the AGGREGATE operator is shown in figure
\ref{Fig.13}. The batch service time obtained in the training
process is a 53-stage PH distribution with an average of
$0.6121(ms)$. In fact, it is similar to a uniform distribution.
Uniform distribution is a special case of PH distribution and can be
fitted via increasing the stage number of PH distribution. The
distribution can be noted as formula (\ref{Formula.09}).

\begin{footnotesize}
\begin{equation}\label{Formula.09}
    \alpha = [\underbrace{0.9992, 0, \cdots, 0.0008}_{53}] \quad T = \left(
                                                                 \begin{array}{cccccc}
                                                                   -72081.18 & 72081.18 & 0 & \cdots & \cdots  & \cdots  \\
                                                                   0 & -72081.18 & 0 & 0 & \cdots  & \cdots \\
                                                                   \vdots & 0 &  \vdots & 72081.18 & 0 & 0 \\
                                                                    & \vdots & \cdots &  -74322.59 & 21604.73 & 72153.12\\
                                                                    &  & \vdots & 0 & -69767.76 & 69767.76 \\
                                                                    &  &  & \vdots & 0 & -29965.91 \\
                                                                 \end{array}
                                                               \right)
\end{equation}
\end{footnotesize}

In conclusion, we obtained a queuing model with a 2-stage PH
distribution input process, a 53-stage distribution service time, a
buffer with fixed size 60, a window (batch) size of 20, one service
desk and FCFS rule. It can be noted as $PH_{2}/PH_{53}^{[20, 20]}
/1/60/\infty/FCFS $. As shown in table \ref{Tab.10}, the major
performance indicators can be obtained according to the $MAP/G^{[a,
b]}/1/N$\cite{elsarticle-num:gupta_analysis_2001} model.

Table \ref{Tab.10} says the predicted error of residence time is
20.6829\% while the queue length error is 7.2083\%¡£Obviously, the
residence time error is unacceptable. We found that the tuple
transmission delay on local area network can't be ignored when it is
compared to the processing time in streaming system. In fact, the
average interval time at the sending end is $0.9457(ms)$ and
increases to $1.1049(ms)$ at the receiving end. By using the tuple
arriving interval sequence at the receiving end, the residence time
is predicted to be $10.5275(ms)$ which reduce the error from
20.6829\% to 7.3683\% and little is changed for the queue length
prediction.

\begin{table}[th]
  \centering
  \caption{Behavior indicators of $PH_{2}/PH_{53}^{[20, 20]}/1/60$ model}\label{Tab.10}
  \small
  \begin{tabular}{ccccc}
  \toprule
  \multicolumn{4}{l}{\textbf{Prediction:}}\\
  \midrule
  Queue Length (L) &  9.5324  &  Waiting Queue Length (Lq)  & 9.500\\
  Residence Time (W) & 9.0134 &   Waiting Time (Wq)  & 8.9837 \\
  Pbusy  & 0.00324  &  Ploss\ & 0.000044 \\
  \multicolumn{4}{l}{\textbf{Observation:}}\\
  Queue Length (L) &   10.2797  &  Error &   20.6829\%\\
  Residence Time (W) &   11.3649 &    Error &   7.2083\% \\
  \bottomrule
  \end{tabular}
\end{table}

It is necessary to explain the observation of residence time of an
output tuple and the queue length. In a batch processing, the tuples
are divided into different groups according to their keys and each
group will generate only one corresponding tuple. Its residence time
is the sum of the average waiting time of the tuples in that group
and the processing time of that group. The queue length is obtained
through recording the tuple number in the whole system (including
queue and operator box) at each tuple arrival and then averaging them.

Here the queue length $N = 60$ and window size $K = 20$ are
relatively small because of the computation limitation of
high-dimensional matrix. An $n$-stage PH distribution has an $n$
dimension transfer matrix. Taking the queue length and service time
distribution into consideration, the dimension of the intermediate
result will reach $2 \times 60 \times 53 = 6360$ here. If the queue
length $N$ reaches 200, it will bring trouble to the computation.

In order to make our model more practical, we obtained numerical
functional relationship between predicted queue length, residence
time and window size. Assuming K at between 10 and 80 $(N = 2K <
200)$, we obtained the functional relationship as expression
(\ref{Formula.10}) :

\begin{equation}\label{Formula.10}
    L = 0.4994K - 0.4526 \qquad W = 0.5518K - 0.5002
\end{equation}

When K = 8000, the prediction value of queue length and residence
time are 3994.4747 tuples and $4.4139(s)$ respectively. In practice,
the observed values are 4008.9335 tuples and $4.9367(s)$ which
indicate the error is 0.3600\% and 10.1861\%. We consider the
functional relation to be reasonable.

\subsection{Queuing Model for Small Window Array AGGREGATE Operator}
\begin{figure}
  \centering
  \includegraphics[width=0.80\textwidth, scale=0.5]{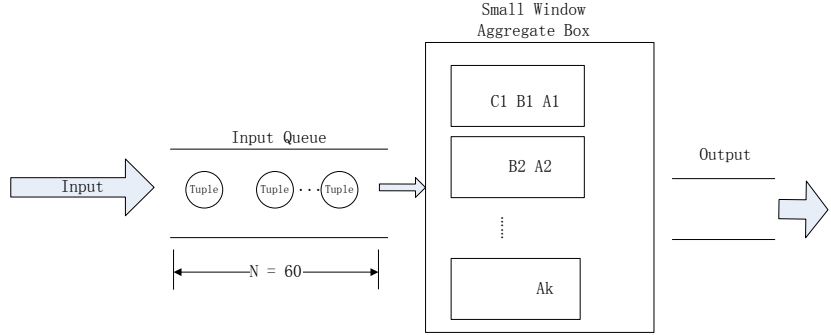}\\
  \caption{Architecture of Small window array Aggregate operator}\label{Fig.14}
\end{figure}

The architecture of the small window array aggregate operator is
shown in figure \ref{Fig.14}. At any time there is a group of small
windows coexisting in the operator box. Each small window that
collects related invocations represents a service (page) instance.
In the case, we take the arrival of a tuple as the opening of a new
small window. In other words, the input of the special aggregate
operator is an instance-level stream. It should be noted that our
aggregate operator locates after the union operator in the
processing network as shown in figure \ref{Fig.07} and thus the
observed input stream is an invocation-level stream with the same
distribution of expression (\ref{Formula.06}). It is necessary to
extract the primary invocations interval in the observed stream and
estimate its distribution. In fact, the instance-level stream is a
2-stage PH distribution as recorded in expression (\ref{Formula.05})
and has an average interval of $9.7780(ms)$.

In our experiment, we got an average service time of $20.7137(s)$ in
the training process by setting window size to be 13 and taking
timeout as $22(s)$. Because of the timeout mechanism, it is
difficult to fitting the service time distribution and hence we only
obtained its average and standard deviation.

The service mode of this AGGREGATE operator is a little interesting.
Each small window in the system will generate \textbf{one} output
tuple. For that tuple, we consider the service begin at the point
when a window is opened (the first invocation tuple is inserted into
it immediately). When the window is closed and the output tuple is
generated the service process ends. It is clear that multi-windows
coexist in the system and thus the service desk works in a
concurrent mode. In other words, the number of service desk is
$C(>1)$ instead of \textbf{one}.

A queuing system can keep in equilibrium state only on condition
that the traffic intensity $ \rho \leqslant 1$, i.e. the
average input rate $\lambda$ should be slower than the process rate
$\mu$. Here we need to adjust service desk number $C$ to make sure
the system stay in a stable state:

\begin{equation}\label{Formula.11}
    \rho = \frac{\lambda}{c\mu}<1\Rightarrow
    c>\frac{\lambda}{\mu}=\frac{1/0.00977780}{1/20.7137}=2118.3984\approx2119
\end{equation}

We can obtain the minimum of C as 2119 in the case, i.e.
there should be at least 2119 small windows in the aggregate
operator box to guarantee the processing capacity. Having the number
of windows and the window size, we can obtain the necessary storage
space of aggregate operator as follows:

\begin{footnotesize}
\begin{equation}\label{Formula.12}
    Numberof(Window)\times Capacityof(Window) \times Sizeof(Tuple) =
    C \times K \times 135 = 2119 \times 13 \times 135 = 3.5466(MB)
\end{equation}
\end{footnotesize}

A conservative estimate suggests that the storage space allocated to
the aggregate operator box is about 32 MB in on a computation node
with 2000MB memory space in total (we observed a minimum memory
consumption via deploying the single AGGREGATE operator on a
computation node), i.e. the operator box can accommodate at least 1,
9119 small windows with the fixed size of 13. The number is much big
than 2119 and hence we can hold the opinion that the service desk is
infinite. As for the queue length, we can configure it through
AURORA\_PAGE and AURORA\_PAGE\_SIZE and hence we can make it to be
long enough. Finally we obtained a model which can be noted as
$G/G/(C\gg2119)$ and gained the indicators through Queuing ToolPak
\cite{elsarticle-num:QTP} by using the average as well as the stand
deviation of both input process and service time.

PH distribution is no longer suitable in this case and thus the
$MAP/G^{[a, b]}/1/N$ model do little here for the multiple service
desk models. Table \ref{Tab.11} predicts the queue length (i.e. the
number of small windows) to be 2120 and observes 1861 in fact. The
difference leads to an error of 12.0251\%. When it comes to
residence time, the error is $(20.7137 - 19.0129) / 20.7137 =
9.04\%$.

\begin{table}[th]
  \centering
  \caption{Behavior indicators of $G/G/10000$ model}\label{Tab.11}
  \begin{threeparttable}
  \small
  \begin{tabular}{cccc}
  \toprule
  \multicolumn{4}{l}{\textbf{Prediction:}}\\
  \midrule
  Queue Length (L)  &  2119.2972  &  Waiting Queue Length (Lq) & $3.1560\times10^{-53}$\\
  Residence Time (W) & 20.7137 &   Waiting Time (Wq)  & 0 \\
  Pwait  & $7.9272\times10^{-5}$  &  Ploss  & 0 \\
  \multicolumn{4}{l}{\textbf{Observation:}}\\
  $\lambda$ \tnote{a} &  102.2704 & $\mu$ \tnote{b} & 0.0483 \\
  Queue Length (L) &   1860.6358  &  Error &   12.0251\% \\
  Residence Time (W) &  19.0129 &    Error &   9.0402\% \\
  \bottomrule
  \end{tabular}
  \begin{tablenotes}
  \item [a] {Input rate }
  \item [b] {Service rate}
  \end{tablenotes}
  \end{threeparttable}
\end{table}

As discussed before, the transmission delay contributes greatly to
the gross error of queue length prediction. The average arriving
interval time adds up to $11.8092(ms)$ in the receiving end but only
$9.7780(ms)$ at the sending end. Calculating the performance
indicators again by using the interval distribution at the receiving
end, we acquired reasonable results. The calculated queue length is
1754.0605 and the error reduces from 12.0251\% to 5.6989\%.

\subsection{Comparison between the AGGREGATE Operators}
Here we conduct a comprehensive comparison between these two window
mechanisms. For small window array approach, the window parameters
are taken as $13/13/22$. On the other hand, the window size and
advance step of normal sliding window are both set to 32000. The
association keys are\textbf{\emph{ service structure, instance
status information (timestamp) and client information}}. The results
are shown in table \ref{Tab.12}.

\begin{table}[th]
  \centering
  \caption{Comprehensive comparison between different window mechanisms}\label{Tab.12}
  \small
  \begin{tabular}{ccc}
    \toprule
    Indicators & Small Window Array & Sliding Widnow \\
    \midrule
    Window Parameters & 13/13/22 & 32000/32000 \\
    Integration Completeness($\gamma=1$) &  80.4815\%  &  72.1083\%\\
    Integration Completeness($\gamma=0.85$) &   90.2979\%  & 73.7587\% \\
    Integration Completeness($\gamma=0.75$) &   91.6553\%  & 74.7302\% \\
    Integration Completeness($0<\gamma\leqslant1$) & 94.0649\% & 76.2744\% \\
    Average Queue Length (L.avg) &   1861(window)  &  15374(tuple) \\
    Average Storage (S.avg) & 3.1148MB & 5.2909MB \\
    Max Queue Length (L.max) & 2007(window) & 32096(tuple) \\
    Max Storage (S.max)& 3.3591MB &5.3910MB \\
    Residence Time (W)  & 19.0192(S) & 16.8402(S) \\
    \bottomrule
  \end{tabular}
\end{table}

Through the quantitative analysis of the two mechanisms, it is clear
to see the advantages of small window array in stateful stream under
service environment. It can not only guarantee high information
integration completeness but also has acceptable performance as well
as resource consumption indicators. Although the sliding window has
approximate completeness when window size $K = 32000$, it not only
consumes more memory but also leads to unstable integration
completeness as tuple arrival pattern varies. We obtained the
completeness here after multiple attempts which is impossible under
real stream scenario. That is to say sliding window is not practical
in stateful stream.

An important note about the sliding window is the computation of
storage space. Aggregate operator will partition the tuples into
different groups according the specified key. We observed that the
maximum, average, minimum group numbers are 3221, 3162 and 2962
respectively. Each group is allocated with the same size 13 which is
deduced in the small window array mechanism. Having the group
number, group size and the tuple size, we can acquire the storage
space easily.

\section{Conclusion and Future work}
In this paper we mainly discussed the stateful stream in service
intelligence context. Thorn problems arise because of the complicate
structure and multi-copy of service on a platform,. We focus on
solving association accuracy, integration completeness and
performance evaluation during the distributed stream information
integration process.

We (1) proposed associate strategies for decentralized information
systematically and conducted comparative analysis in detail (2)
analyzed a new mechanism of small window array and employed the
cumulative distribution function approach to guide the setting of
window parameters which can guarantee completeness of integration
(3) built reasonable queuing models for the typical streaming
operators involved in the paper to predict performance indexes.
Experiments show that our work is convincing and effective.

There are also some meaningful issues that are in need of further
study:
\begin{enumerate}
  \item Building queuing model for JOIN operator. Generally the join
operator accepts two input streams and connects the related tuples
according to specified keys. Independent buffers will be reserved
for each stream. To construct sound model for JOIN operator, it is
necessary to obtain a deep understanding about the working
principles and draw support from the embedded markov chain,
supplementary variable methods.
  \item How to evaluate the goodness of the specified completeness
  factor $\alpha$? In sec.\ref{Sec.Integration}, we simply give a
  value subjectively. Does the factor has something to do with the
  users' error tolerance? How can we establish appropriate mapping
  between error tolerance and completeness index?
  \item  Intuitively, the probability a tuple will be dropped at its arriving given by the
  queuing model can contribute to the load shielding of stream engine to some
  degree. Is this approach really practical?
  \item Data stream network optimization. Though queuing theory is a
powerful tool in behavior analysis when comparing with the cost
model \cite{elsarticle-num:xing_providing_2006}, it also performs
pool on how to construct an optimized operator network. Directed
Acyclic Graph theory, Type theory and multi-query optimization of
database may be of great significance in this sphere.
\end{enumerate}


\bibliographystyle{elsarticle-num}
\bibliography{Streaming}

\begin{thebibliography}{10}
\expandafter\ifx\csname url\endcsname\relax
  \def\url#1{\texttt{#1}}\fi
\expandafter\ifx\csname urlprefix\endcsname\relax\def\urlprefix{URL }\fi
\expandafter\ifx\csname href\endcsname\relax
  \def\href#1#2{#2} \def\path#1{#1}\fi

\bibitem{elsarticle-num:Erl_SOAPrinciples}
T.~Erl, SOA: principles of service design, Prentice Hall Press, Upper Saddle
  River, NJ, USA, 2007.

\bibitem{elsarticle-num:jones_toward_2005}
S.~Jones, Toward an acceptable definition of service, {IEEE} software 22~(1)
  (2005) 87--93.

\bibitem{elsarticle-num:software_service}
Software as a service, {http://en.wikipedia.org/wiki/Software\_as\_a\_service}.

\bibitem{elsarticle-num:li_adaptive_2008}
S.~M. Li, C.~Ding, C.~H. Chi, J.~Deng, Adaptive quality recommendation
  mechanism for software service provisioning, in: Proceedings of the 2008
  {IEEE} international conference on web services, {IEEE}, 2008, pp. 169--176.

\bibitem{elsarticle-num:keller_defining_SLA_2002}
A.~Keller, H.~Ludwig, Defining and monitoring service level agreements for
  dynamic {e-Business}, 2002.

\bibitem{elsarticle-num:sarwar_item-based_2001}
B.~Sarwar, G.~Karypis, J.~Konstan, J.~Reidl, Item-based collaborative filtering
  recommendation algorithms, in: Proceedings of the 10th international
  conference on World Wide Web, {ACM}, 2001, pp. 285--295.

\bibitem{elsarticle-num:burtch_user-generated_content}
G.~Burtch, Y.~Hong, {User-Generated Content (UGC): Developing a Theoretical
  Framework to Inform is Research}, SSRN eLibrary.

\bibitem{elsarticle-num:schmidt_quality--service-aware_2007}
S.~Schmidt, Quality-of-service-aware data stream processing, Ph.D. thesis,
  Dresden University of Technology (2006).

\bibitem{elsarticle-num:koudas_data_2003}
N.~Koudas, D.~Srivastava, Data stream query processing, in: Proceedings of the
  4th international conference on web information systems engineering, {IEEE},
  2003, p. 374.

\bibitem{elsarticle-num:babcock_models_2002}
B.~Babcock, S.~Babu, M.~Datar, R.~Motwani, J.~Widom, Models and issues in data
  stream systems, in: Proceedings of the 21st ACM SIGMOD-SIGACT-SIGART
  symposium on principles of database systems, {ACM}, 2002, pp. 1--16.

\bibitem{elsarticle-num:golab_issues_2003}
L.~Golab, M.~T. \"{?}zsu, Issues in data stream management, {ACM} Sigmod Record
  32~(2) (2003) 5--14.

\bibitem{elsarticle-num:babcock_distributed_2003}
B.~Babcock, C.~Olston, Distributed top-k monitoring, in: Proceedings of the
  2003 {ACM} {SIGMOD} international conference on management of data, {ACM},
  2003, pp. 28--39.

\bibitem{elsarticle-num:olston_adaptive_2003}
C.~Olston, J.~Jiang, J.~Widom, Adaptive filters for continuous queries over
  distributed data streams, in: Proceedings of the 2003 {ACM} {SIGMOD}
  international conference on management of data, {ACM}, 2003, pp. 563--574.

\bibitem{elsarticle-num:madden_fjording_2002}
S.~Madden, M.~J. Franklin, Fjording the stream: An architecture for queries
  over streaming sensor data, in: Proceedings of the 18th international
  conference on data engineering, {IEEE}, 2002, pp. 555--566.

\bibitem{elsarticle-num:borealis}
The borealis project, \url{http://www.cs.brown.edu/research/borealis/public/}.

\bibitem{elsarticle-num:abadi_design_2005}
D.~J. Abadi, Y.~Ahmad, M.~Balazinska, M.~Cherniack, J.~hyon Hwang, W.~Lindner,
  A.~S. Maskey, E.~Rasin, E.~Ryvkina, N.~Tatbul, Y.~Xing, S.~Zdonik, The design
  of the borealis stream processing engine, in: Proceedings of the 2nd biennial
  conference on innovative data systems research, Citeseer, Asilomar, CA, USA,
  2005, pp. 277--289.

\bibitem{elsarticle-num:TelegraphCQ}
Telegraphcq, \url{http://telegraph.cs.berkeley.edu/telegraphcq/v2.1/}.

\bibitem{elsarticle-num:TelegraphCQ_Uncertain_World}
S.~Chandrasekaran, O.~Cooper, A.~Deshpande, M.~J. Franklin, J.~M. Hellerstein,
  W.~Hong, S.~Krishnamurthy, S.~Madden, V.~Raman, F.~Reiss, M.~A. Shah,
  Telegraphcq: Continuous dataflow processing for an uncertain world, in:
  Proceedings of the 1st biennial conference on innovative data systems
  research, 2003.

\bibitem{elsarticle-num:TelegraphCQ_Adaptive_Query}
S.~Madden, M.~Shah, J.~M. Hellerstein, V.~Raman, Continuously adaptive
  continuous queries over streams, in: Proceedings of the 2002 {ACM} {SIGMOD}
  international conference on management of data, {ACM}, 2002, pp. 49--60.

\bibitem{elsarticle-num:Stanford_STREAM_Main}
Stanford stream data manager, \url{http://infolab.stanford.edu/stream/}.

\bibitem{elsarticle-num:Stanford_DSMS}
A.~Arasu, B.~Babcock, S.~Babu, J.~Cieslewicz, M.~Datar, K.~Ito, R.~Motwani,
  U.~Srivastava, J.~Widom, Stream: The stanford data stream management system,
  Technical Report 2004-20, Stanford InfoLab (2004).

\bibitem{elsarticle-num:Borealis_Guide}
B.~Team, Borealis application programmer's guide, Technical report, Brandeis
  University, Brown University, MIT (2006).

\bibitem{elsarticle-num:gross_fundamentals_2008}
D.~Gross, Fundamentals of queueing theory, {Wiley-India}, 2008.

\bibitem{elsarticle-num:osogami_analysis_2005}
T.~Osogami, Analysis of multi-server systems via dimensionality reduction of
  markov chains, Ph.D. thesis, Carnegie Mellon University (2005).

\bibitem{elsarticle-num:Erlang_distribution}
Erlang distribution, \url{http://en.wikipedia.org/wiki/Erlang\_distribution}.

\bibitem{elsarticle-num:xu_performance_2008}
X.~Xu, W.~Wang, S.~Xu, Performance of a queuing model with {Hyper-Erlang}
  distribution service for wireless network nodes, in: Proceedings of the
  {IEEE} international conference on wireless communications networking and
  mobile computing, {IEEE}, 2008, pp. 1--4.

\bibitem{elsarticle-num:phase-type_distribution}
Phase-type distribution,
  \url{http://en.wikipedia.org/wiki/Phase-type\_distribution}.

\bibitem{elsarticle-num:PH_EM}
A.~Th¨¹mmler, P.~Buchholz, M.~Telek, A novel approach for phasetype fitting
  with the em algorithm, IEEE Transactions on Dependable and Secure Computing 3
  (2006) 245--258.

\bibitem{elsarticle-num:gupta_analysis_2001}
U.~C. Gupta, P.~V. Laxmi, Analysis of the {$MAP/G^{[a, b]}/1/N$} queue,
  Queueing systems 38~(2) (2001) 109--124.

\bibitem{elsarticle-num:banik_finite-buffer_2009}
A.~D. Banik, U.~C. Gupta, M.~L. Chaudhry, Finite-buffer bulk service queue
  under markovian service process: {$GI/MSP^{[a, b]}/1/N$}, Stochastic Analysis
  and Applications 27~(3) (2009) 500--522.

\bibitem{elsarticle-num:laxmi_finite-buffer_1999}
P.~V. Laxmi, U.~C. Gupta, On the finite-buffer bulk-service queue with general
  independent arrivals: {$GI/M^{[b]}/1/N$}, Operations Research Letters 25~(5)
  (1999) 241--245.

\bibitem{elsarticle-num:SOA_Reliability}
V.~Cortellessa, V.~Grassi, Reliability modeling and analysis of
  service-oriented architectures, Test and Analysis of Web Services (2007)
  339--362.

\bibitem{elsarticle-num:htmlunit}
Htmlunit, \url{http://htmlunit.sourceforge.net/}.

\bibitem{elsarticle-num:Wiki_Trace}
G.~Urdaneta, G.~Pierre, M.~van Steen, Wikipedia workload analysis for
  decentralized hosting, Elsevier Computer Networks 53~(11) (2009) 1830--1845.

\bibitem{elsarticle-num:wikimedia}
M.~Bergsma, Wikimedia architecture,
  \url{http://www.nedworks.org/~mark/presentations/san/Wikimedia\%20architecture.pdf}.

\bibitem{elsarticle-num:EM_MAP}
P.~Buchholz, An {EM-algorithm} for {MAP} fitting from real traffic data,
  Computer Performance (2003) 218--236.

\bibitem{elsarticle-num:QTP}
Queueing {ToolPak} 4.0,
  \url{http://apps.business.ualberta.ca/aingolfsson/qtp/}.

\bibitem{elsarticle-num:xing_providing_2006}
Y.~Xing, J.~H. Hwang, U.~Cetintemel, S.~Zdonik, Providing resiliency to load
  variations in distributed stream processing, in: Proceedings of the 32nd
  international conference on very large data bases, {VLDB} Endowment, 2006,
  pp. 775--786.

\end{thebibliography}
\end{document}